\documentclass[fleqn,usenatbib]{mnras}

\usepackage[T1]{fontenc}
\usepackage{ae,aecompl}

\usepackage{graphicx}	
\usepackage{amsmath}	
\usepackage{amssymb}	

\title[Gaia-unWISE AGN catalogues]{Catalogues of Active Galactic Nuclei From Gaia and unWISE Data}

\author[Yiping Shu et al.]{Yiping Shu,$^{1}$\thanks{E-mail: yiping.shu@ast.cam.ac.uk,nwe@ast.cam.ac.uk}\thanks{Royal Society -- K. C. Wong International Fellow}
Sergey E. Koposov,$^{2, 1}$
N. Wyn Evans,$^{1}$
Vasily Belokurov,$^{1}$
\newauthor
Richard G. McMahon,$^{1, 3}$
Matthew W. Auger$^{1, 3}$
and Cameron A. Lemon$^{1, 3}$
\\
$^{1}$ Institute of Astronomy, Madingley Rd, Cambridge, CB3 0HA, UK\\
$^{2}$ McWilliams Center for Cosmology, Carnegie Mellon University, 5000 Forbes Ave, Pittsburgh, PA 15213, USA\\
$^{3}$ Kavli Institute for Cosmology, University of Cambridge, Madingley Road, Cambridge, CB3 0HA, UK
}

\date{Accepted XXX. Received YYY; in original form ZZZ}

\pubyear{2019}

\begin{document}
\label{firstpage}
\pagerange{\pageref{firstpage}--\pageref{lastpage}}
\maketitle

\begin{abstract}
We present two catalogues of active galactic nucleus (AGN) candidates selected from the latest data of two all-sky surveys -- Data Release 2 (DR2) of the \emph{Gaia} mission and the unWISE catalogue of the \emph{Wide-field Infrared Survey Explorer} (\emph{WISE}). We train a random forest classifier to predict the probability of each source in the \emph{Gaia}-unWISE joint sample being an AGN, $P_{\rm RF}$, based on \emph{Gaia} astrometric and photometric measurements and unWISE photometry. The two catalogues, which we designate C75 and R85, are constructed by applying different $P_{\rm RF}$ threshold cuts to achieve an overall completeness of 75\% ($\approx$90\% at \emph{Gaia} $G\leq20$ mag) and reliability of 85\% respectively. The C75 (R85) catalogue contains 2,734,464 (2,182,193) AGN candidates across the effective 36,000 deg$^2$ sky, of which $\approx$0.91 (0.52) million are new discoveries. Photometric redshifts of the AGN candidates are derived by a random forest regressor using \emph{Gaia} and \emph{WISE} magnitudes and colours. The estimated overall photometric redshift accuracy is 0.11. Cross-matching the AGN candidates with a sample of known bright cluster galaxies, we identify a high-probability strongly-lensed AGN candidate system, SDSS\,J1326$+$4806, with a large image separation of 21\farcs06. 
All the AGN candidates in our catalogues will have $\sim$5-year long light curves from \emph{Gaia} by the end of the mission, and thus will be a great resource for AGN variability studies. Our AGN catalogues will also be helpful in AGN target selections for future spectroscopic surveys, especially ones in the southern hemisphere. The C75 catalogue can be downloaded at \url{https://www.ast.cam.ac.uk/~ypshu/AGN_Catalogues.html}
\end{abstract}




\section{Introduction}

Active galactic nuclei (AGNs) are compact cores in active galaxies that emit strong electromagnetic radiation over a broad wavelength range. They are believed to be powered by the accretion activities of the central supermassive black holes \citep[e.g.,][]{Lynden-Bell69, Rees84, Tanaka95}. Very luminous AGNs can also be referred to as quasars (also known as QSOs). Large samples of AGNs are of great importance in astrophysics. They can be used to define celestial reference frames \citep[e.g.,][]{Ma98, Fey15, Mignard16, Mignard18}. The variability from AGNs has been used to constrain the properties of supermassive black holes and the fuelling mechanisms \citep[e.g.,][]{Blandford82, Vanden04, Liu08, Li08, MacLeod10, Shen15, LaMassa15, Yang18}. Among the most luminous sources in the sky, AGNs have been detected back to within the first billion years of the Universe and help to understand the growth of supermassive black holes \citep[e.g.,][]{Fan06, Wu15, Wang18, Pons19, Shen19}. In addition, AGNs have been suggested to play an important role in regulating the formation and evolution of host galaxies \citep[e.g.,][]{Silk98, Kang06, Fabian12, Dubois13}. Furthermore, spectroscopic observations of AGNs across a wide redshift range can probe the neutral hydrogen fraction in the intergalactic medium and mass distribution in general, which further constrain the history of reionization and cosmological parameters \citep[e.g.,][]{Mortlock11, Delubac15, Bautista17, Banados18, Zhao19}.

AGNs can be selected based on X-ray observations or by ultraviolet (UV), infrared (IR) or optical photometry and spectroscopy. Each has different biases that affect the resulting samples. Optical identification militates against heavily obscured AGNs, whilst X-ray selected samples are more robust against obscuration. Mid-IR and optical identification can be hampered by the host galaxy's emission, and this is known to bias against AGNs accreting at low fractions of the Eddington limit. Mid-IR and X-ray observations are usually space-based because of the Earth's atmosphere, though the latter require significantly longer exposure time.

The advent of data from the \emph{Wide-field Infrared Survey Explorer} (\emph{WISE}) \citep{Wright10} spurred the construction of AGN catalogues based solely on mid-IR data. The \emph{WISE} mission imaged the entire sky in four mid-IR bands, centred at 3.4, 4.6, 12, and 22 $\mu$m, referred to as $W1$, $W2$, $W3$, and $W4$, respectively. As noticed in previous work \citep[e.g.,][]{Lacy04, Stern05, Stern12, Nikutta14}, AGNs tend to have redder $W1-W2$ colours relative to stars and inactive, low-redshift galaxies. As a result, a number of work relied on the $W$1-$W$2 colour in selecting AGNs from the AllWISE Data Release \citep[e.g.,][]{Assef13, Secrest15, Assef18}. Very recently, \citet{Schlafly19} provided an enhanced unWISE catalogue of roughly 2.03 billion objects that is based on significantly deeper imaging from use of coadds of all publicly available \emph{WISE} data \citep{Lang14, Meisner17a, Meisner17b} and has a superior treatment of crowding. This paper provides the first AGN catalogues using the unWISE data.

Nevertheless, the mid-IR-only selection techniques have some limitations. The first is the generally poor imaging resolution of mid-IR data ($\sim 6^{\prime \prime}$ in \emph{WISE} $W1$ and $W2$ bands). As a result, source blending can become a considerable issue and lead to mis-classifications or render the blended data unusable. Secondly, some non-AGNs have similarly red $W1-W2$ colours as AGNs, which are difficult to distinguish with mid-IR data alone. For example, high-redshift ($z \gtrsim 1.2$) early-type galaxies also have red $W1-W2$ colours because of the rest-frame 1.6 $\mu$m stellar bump being shifted beyond the $W1$ band at $z \gtrsim 1.2$ \citep[e.g.,][]{Papovich08, Papovich10, Galametz12, Yan13}. This type of contamination is not significant in the AllWISE data because the characteristic magnitude of high-redshift early-type galaxies in the $W2$ band is about 16.7 mag \citep[e.g.,][]{Mancone10}, at which the AllWISE catalogue is very incomplete. However, it becomes more pronounced in the deeper unWISE catalogue, which reaches $\approx$50\% complete at $W2 = 16.7$ mag \citep{Schlafly19}. In addition, young stellar objects (YSOs), dusty asymptotic giant branch (AGB) stars, and extended planetary nebulae are also found to have similar $W1-W2$ colours as AGNs \citep[e.g.,][]{Rebull10, Koenig12, Nikutta14, Assef18}. 

Optical data have also been used to efficiently select AGNs, through mostly the ``UV excess'' method or multi-colour cuts \citep[e.g.,][]{Sandage65, Warren87, Hewett95, Richards02, Richards04, Smith05, Schneider10, Bovy11, Myers15}. Furthermore, the combination of optical and IR data is found to improve the success rate of AGN selections \citep[e.g.,][]{Wu10, Maddox12, McGreer13, Richards15, Wang16}. High-redshift galaxies, YSOs, and AGB stars can also be better identified with the inclusion of optical data. Hitherto, the sky coverage has been limited due to the lack of an all-sky optical survey. However, the European Space Agency's \emph{Gaia} space telescope, launched in 2013, is delivering precise astrometry and optical photometry for more than a billion sources across the entire sky for the first time \citep{Prusti16}. \emph{Gaia} measures three broadband photometry \citep{Evans18}, i.e. $G$ band (330--1050 nm), the blue prism photometer (BP, 330--680 nm), and the red prism photometer (RP, 630--1050 nm). On the 25$^{\mathrm{th}}$ April 2018, \emph{Gaia} delivered its second data release \citep[Gaia DR2,][]{Brown18} containing astrometry and photometry for 1.69 billion sources, based on the first 22 months of operation. 

In this paper, we construct new all-sky AGN catalogues based on the combination of these two latest catalogues from \emph{Gaia} and unWISE. This paper is organised as follows. Section~\ref{sect:sample} describes some properties of the \emph{Gaia}-unWISE sample. Section~\ref{sect:method} explains the methods and procedures used to classify AGNs and estimate their photometric redshifts. Section~\ref{sect:results} presents our catalogues of AGN candidates. Discussions and conclusion are given in Sections~\ref{sect:discussions} and \ref{sect:conclusion}. Throughout the paper, we adopt a cosmological model with $\Omega_m = 0.308$, $\Omega_{\Lambda} = 0.692$, and $H_0 = 67.8$ km s$^{-1}$ Mpc$^{-1}$ \citep{Planck15}. All the magnitudes are given in the Vega system, unless otherwise noted. 

\begin{figure*}
    \centering
    \includegraphics[width=0.48\textwidth]{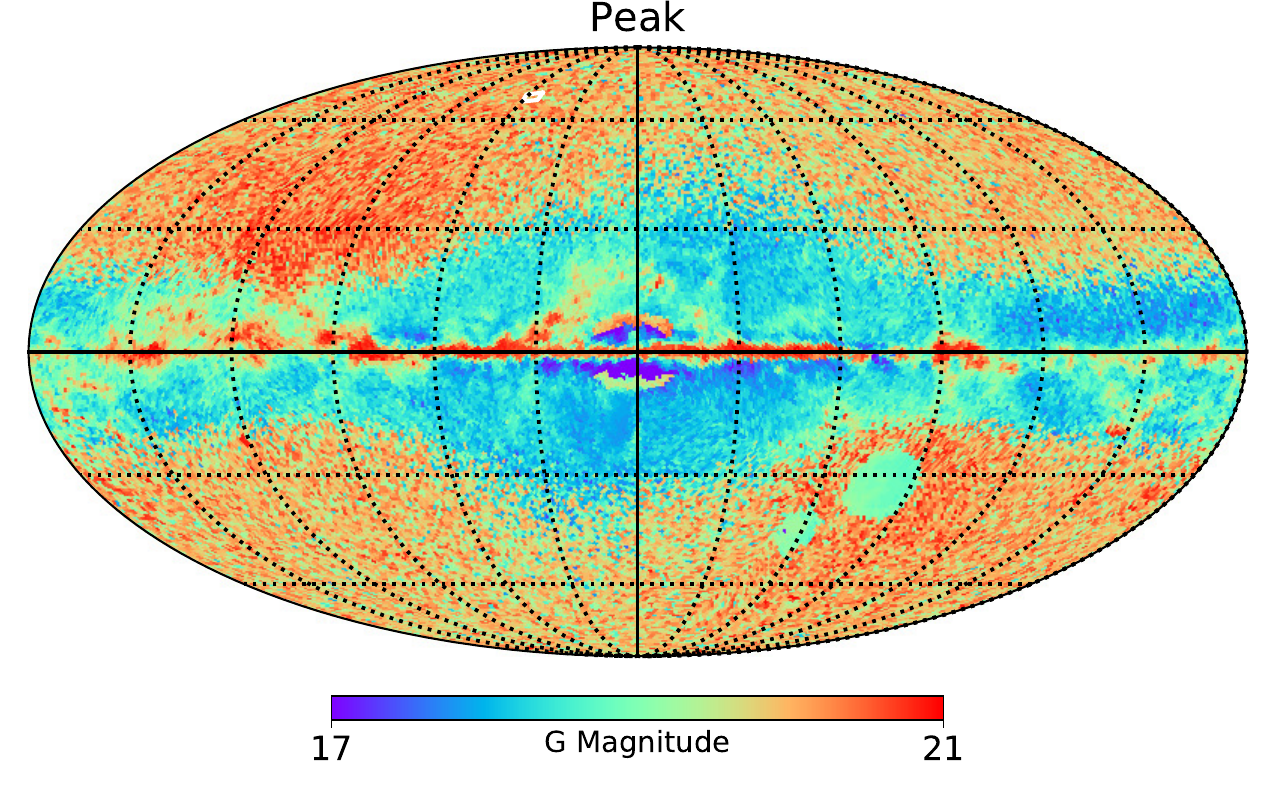}
    \includegraphics[width=0.48\textwidth]{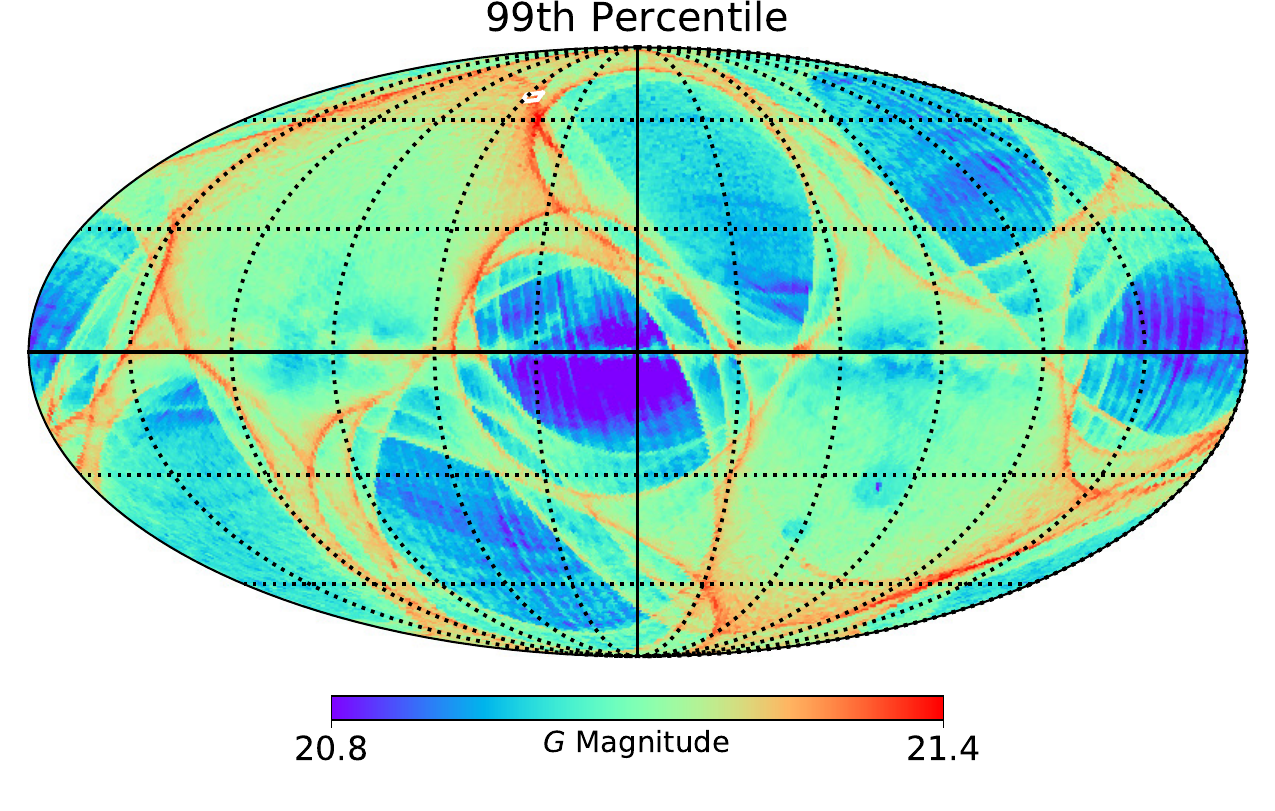}
    \includegraphics[width=0.42\textwidth]{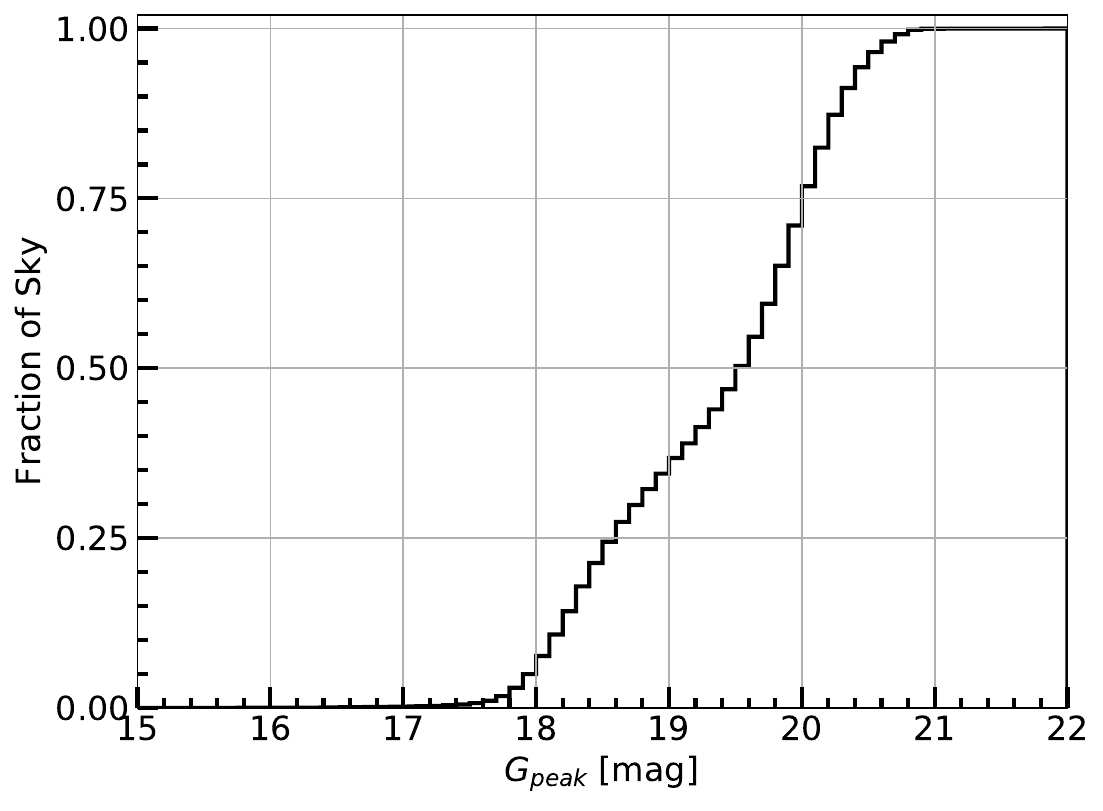}
    \hspace{0.8cm}
    \includegraphics[width=0.42\textwidth]{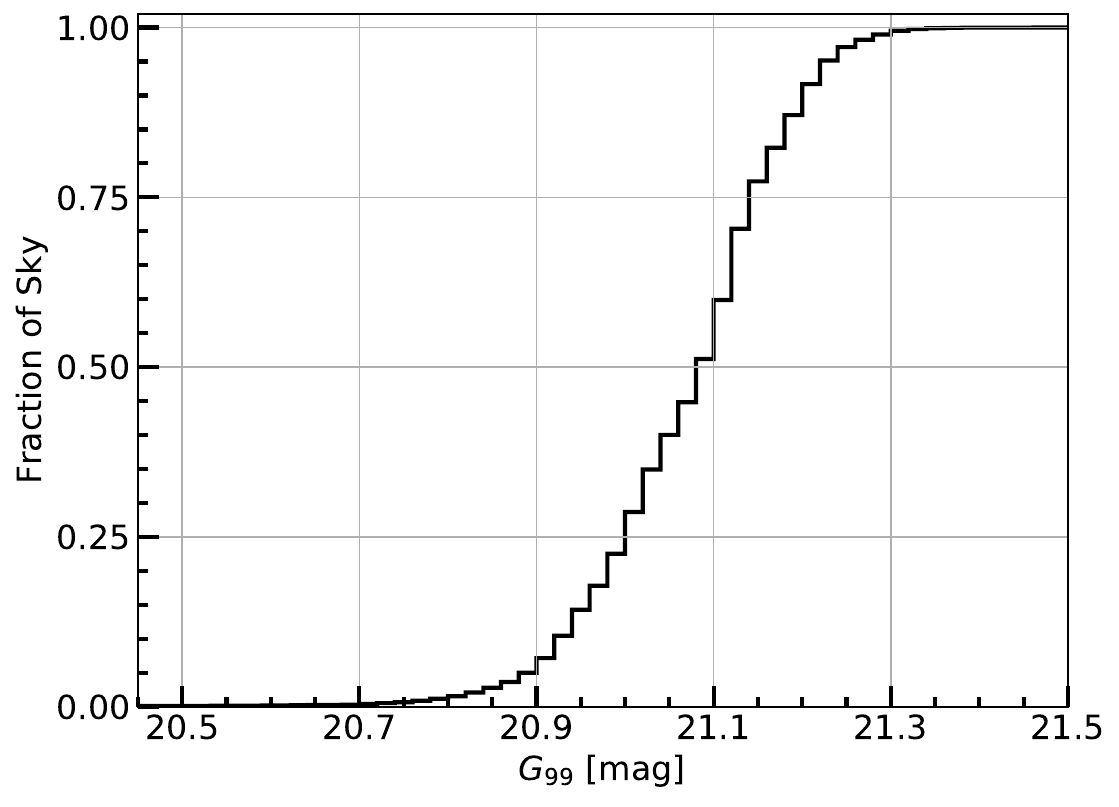}
    \caption{\label{fig:Gaia_G_Limit_Distribution} \emph{Top:} Spatial distributions in Mollweide projection (cell size of $\approx$0.84 deg$^2$) of $G_{\rm peak}$ (\emph{left}) and $G_{\rm 99}$ (\emph{right}) for the \emph{Gaia}-unWISE subsample with $G \geq 16$ mag. The white polygon indicates the location of the Bo\"{o}tes field (at $l \approx 57^{\circ}$, $b \approx 67^{\circ}$). \emph{Bottom:} One-dimensional cumulative sky coverage histograms (bin size of 0.1 mag) of $G_{\rm peak}$ (\emph{left}) and $G_{\rm 99}$ (\emph{right}) for the same \emph{Gaia}-unWISE subsample. }
\end{figure*}

\section{Sample Properties}
\label{sect:sample}

\subsection{Data Preparation}

To build the \emph{Gaia}-unWISE sample for AGN selection, we perform a nearest neighbour cross-match between the \emph{Gaia} DR2 catalogue (the leading catalogue) and the unWISE catalogue using a matching radius of 2$^{\prime \prime}$. In the cross-match process, we only consider sources with non-zero fluxes in both $W1$ and $W2$ bands. As will be shown later, this requirement reduces the number of AGNs in the sample by $\sim 2.6\%$ relative to requiring non-zero flux in $W1$ alone. We take into account the proper motions of sources (as provided by \emph{Gaia} DR2) in the cross-match process because the source positions in the \emph{Gaia} DR2 catalogue and the unWISE catalogue are given at different reference epochs. The \emph{Gaia}-unWISE sample thus includes 641,266,363 unique \emph{Gaia} sources (corresponding to 564,948,465 unique unWISE sources). One thing to note is that due to the design of the \emph{Gaia} mission, mostly point-like objects can be detected by \emph{Gaia}, so the \emph{Gaia}-unWISE sample consists of stars, AGNs, as well as bright and compact (presumably star-forming) regions in extended galaxies. 

\subsection{Completeness \& Depth of the \emph{Gaia}-unWISE Sample}
\label{sect:completeness}

It is known that the \emph{Gaia} completeness and limiting magnitude exhibit complex spatial variation patterns, primarily related to the \emph{Gaia} scanning law \citep[e.g.][]{Arenou18}. However, the completeness and limiting magnitude for the \emph{Gaia}-unWISE sample is still unclear. We thus compute the peak and 99$th$ percentile in the \emph{Gaia} $G$-band magnitude distribution in individual spatial bins for all the $\approx 567$ million sources in the \emph{Gaia}-unWISE sample with $G \geq 16$ mag. The peak $G$ magnitude, $G_{\rm peak}$, should be a good indicator of the completeness, and the 99$th$ percentile in $G$, $G_{99}$, has been used to quantify the limiting magnitude \citep{Arenou18}. Figure~\ref{fig:Gaia_G_Limit_Distribution} shows the spatial distributions and one-dimensional cumulative sky coverage histograms of $G_{\rm peak}$ and $G_{99}$ for the \emph{Gaia}-unWISE subsample. We point out that although shown in the \emph{Gaia} $G$-band magnitude, these maps and histograms have also taken into account the incompleteness and limiting magnitudes in $W1$ and $W2$ of the unWISE catalogue. In particular, the brighter $G_{\rm peak}$ structures at low latitudes and towards the bulge and Magellanic Clouds are primarily caused by the brighter incompleteness limits in $W1$ and $W2$ (see Figures~\ref{fig:unWISE_W1_Limit_Distribution} and \ref{fig:unWISE_W2_Limit_Distribution}). We also find that the \emph{Gaia}-unWISE sample is complete at $G \approx 19.5$ mag in more than 50\% of the sky. The $G_{99}$ map clearly shows the \emph{Gaia} scanning law, where the limiting magnitude is deeper in regions that have more repeated observations by \emph{Gaia}. This is primarily because faint sources that have more repeated observations by \emph{Gaia} tend to have more precise astrometric and photometric measurements and are more likely to be included in the \emph{Gaia} DR2 catalogue relative to sources in the less \emph{Gaia}-scanned regions. The faintest limiting magnitude of the \emph{Gaia}-unWISE sample is about $G$ = 21.4 mag, and more than 50\% of the sky has a limiting magnitude fainter than $G \approx 21.1$ mag. We note that the overall limiting magnitude and completeness for the \emph{Gaia}-unWISE sample will improve in the near future as more repeated \emph{Gaia} observations will be conducted across the whole sky. 

The completeness of the expected AGNs in the \emph{Gaia}-unWISE sample needs to be assessed separately. Figure~\ref{fig:unWISE_W1_Limit_Distribution} shows that the unWISE catalogue is complete at $W1 \approx 16.5$ mag in more than 90\% of the sky. As will be shown later, the expected AGNs in the \emph{Gaia}-unWISE sample generally have $G-W1 > 3$ mag. Considering the $G_{\rm peak}$ distribution in Figure~\ref{fig:Gaia_G_Limit_Distribution}, it is suggested that the expected AGNs will be complete at $G \simeq 19.5$ mag in more than 50\% of the sky (mostly at high latitudes of $\left | b \right | > 20^{\circ}$).

\subsection{AGN Density in the \emph{Gaia}-unWISE Sample}
\label{sect:number_density}

To estimate the expected AGN number density in the \emph{Gaia}-unWISE sample, we use the deep and wide Bo\"{o}tes field of the NOAO Deep Wide-Field Survey \citep[NDWFS,][]{Jannuzi99}. The Bo\"{o}tes field is a $\sim 9.2$ deg$^2$ region centred at approximately R.A.= 218$^{\circ}$, Decl. = 34$^{\circ}$ (indicated by the small, white polygon in Figure~\ref{fig:Gaia_G_Limit_Distribution}) with deep observations in a broad range of (up to 17) filter bands from UV to mid-IR, and therefore has been used for quantifying the performance of AGN selection techniques and AGN studies in general \citep[e.g.,][]{Assef10, Assef13, Chung14, Assef18, Williams18}. In particular, we make use of the catalogue from \citet{Chung14} that contains 431,038 sources extracted from the Bo\"{o}tes field, referred to as the Bo\"{o}tes source catalogue, down to $R \lesssim 23.9$ mag, which should be complete towards the faint end for our purpose as the limiting magnitude of the \emph{Gaia}-unWISE sample is $G \sim 21$ mag. At the bright end, the typical saturation limit of the NDWFS survey is $R \simeq 17$ mag \citep{Chung14}, which roughly corresponds to $G \approx 17$ mag. For every source in the Bo\"{o}tes source catalogue, \citet{Chung14} fitted its observed spectral energy distribution (SED) with stellar, galaxy, and galaxy$+$AGN spectral templates, based on which one can decide whether the source is a star, a galaxy, or an AGN. 

\begin{figure}
    \centering
    \includegraphics[width=0.48\textwidth]{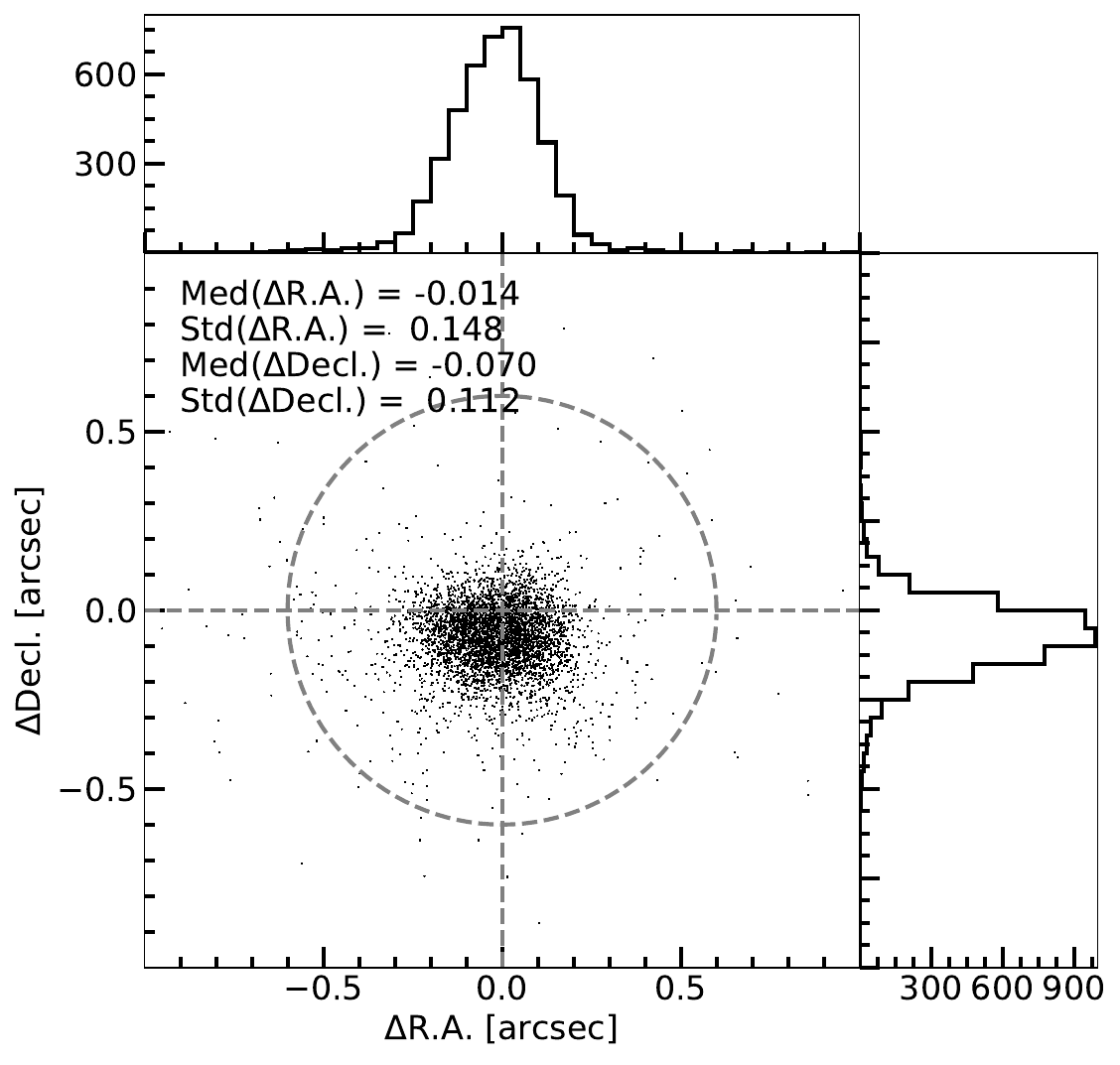}
    \caption{\label{fig:bootes} Positional offsets from \emph{Gaia} DR2 to the Bo\"{o}tes catalogue for the 4,564 matched sources using a matching radius of 1$^{\prime \prime}$. Matches with separations $\leq$0\farcs6 (enclosed by the grey dashed circle) are considered as true matches.}
\end{figure}

We first select a $2^{\circ} \times 2^{\circ}$ sub-region centred at R.A.= 218$^{\circ}$, Decl. = 34$^{\circ}$ from the Bo\"{o}tes field, which contains 159,754 sources from the Bo\"{o}tes source catalogue. We perform a nearest neighbour cross-match between these sources and the \emph{Gaia}-unWISE sample with a matching radius of 1$^{\prime \prime}$. Considering that the source positions are given at different reference epochs, we apply a correction to the \emph{Gaia} DR2 positions in the cross-match process for sources with well-measured proper motions (i.e. S/N > 5), and obtain 4,564 matched sources. The unmatched ones are mostly either extended or fainter sources that are not catalogued in \emph{Gaia} and/or unWISE. Figure~\ref{fig:bootes} shows offsets from positions in \emph{Gaia} DR2 to positions in the Bo\"{o}tes catalogue for the matched 4,564 sources after proper-motion corrections. We find that the median positional offsets are -0\farcs014 in the right ascension direction and -0\farcs07 in the declination direction. More than 99\% (4523) of the matched sources have absolute positional offsets less than 0\farcs6, which are considered as true matches. Further removing sources whose SEDs are better fitted by stellar templates instead of the galaxy$+$AGN templates as indicated by the reduced $\chi^2_{\nu}$ values, i.e. $\chi^2_{\nu} {\rm (star)} \leq \chi^2_{\nu} {\rm (galaxy + AGN)}$, we obtain 718 extragalactic sources in this sub-region. 

To determine how many of the extragalactic sources are AGNs, we consider two metrics that have been previously used for the Bo\"{o}tes source catalogue. The first is the $F$ ratio derived from the reduced $\chi^2_{\nu}$ values and degrees of freedom by \citet{Chung14}. They suggested that a threshold of $F > 10$ should yield a reasonably complete and clean AGN sample. On the other hand, \citet{Assef18} defined a parameter $\hat{a}$, which is the AGN contribution to  the total luminosity based on the SED fitting results, and used $\hat{a} > 0.5$ for selecting AGN candidates. To decide which AGN-selection criterion is appropriate for our purpose, we consider the Sloan Digital Sky Survey (SDSS) DR14 QSO catalogue \citep{Paris18}, based on which our AGN classification is calibrated (as will be shown later). In the $2^{\circ} \times 2^{\circ}$ sub-region, there are 89 DR14 QSOs that are in the \emph{Gaia}-unWISE sample and have been assigned an F ratio and an $\hat{a}$ value by \citet{Chung14}. We find that requiring $F > 10$ or $\hat{a} > 0.5$ alone only recovers 76 or 80 DR14 QSOs, while requiring $(F > 10$ OR $\hat{a} > 0.5)$ can recover 84 DR14 QSOs (i.e. $\approx$95\%). We therefore assume that sources with either $F > 10$ or $\hat{a} > 0.5$ can be considered as AGNs that will be detected in this work.

\begin{table*}
\centering
\caption{\label{tb:features} Features considered in the AGN classification.}
\begin{tabular}{l l}
\hline
\hline
Feature & Description \\
\hline
PLXSIG & parallax significance defined as $\mid \frac{{\tt PARALLAX}}{{\tt PARALLAX\_ERROR}} \mid$, set to -999 if null \\
PMSIG & proper motion significance defined as  $\sqrt{(\frac{{\tt PMRA}}{{\tt PMRA\_ERROR}})^2+(\frac{{\tt PMDEC}}{{\tt PMDEC\_ERROR}})^2}$, set to -999 if null \\
G & Extinction-corrected \emph{Gaia} $G$-band mean magnitude ({\tt PHOT\_G\_MEAN\_MAG}) \\
G\_VAR & Variation in \emph{Gaia} $G$-band flux defined as $\sqrt{{\tt PHOT\_G\_N\_OBS}} \times \frac{{\tt PHOT\_G\_MEAN\_FLUX\_ERROR}}{{\tt PHOT\_G\_MEAN\_FLUX}}$ \\
BP-G & Extinction-corrected \emph{Gaia} BP-$G$ colour ({\tt BP\_G}), set to 999 if null \\
G-RP & Extinction-corrected \emph{Gaia} $G$-RP colour ({\tt G\_RP}), set to 999 if null \\
BPRP & Extinction-corrected \emph{Gaia} BP-RP colour ({\tt BP\_RP}), set to 999 if null \\
BPRP\_EF & BP/RP excess factor ({\tt PHOT\_BP\_RP\_EXCESS\_FACTOR}) \\
AEN & Excess noise of the source ({\tt ASTROMETRIC\_EXCESS\_NOISE}) \\
GOF & Goodness-of-fit statistic of the astrometric solution ({\tt ASTROMETRIC\_GOF\_AL}) \\
CNT1 & Number of \emph{Gaia} sources within a 1$^{\prime \prime}$-radius circular aperture \\
CNT2 & Number of \emph{Gaia} sources within a 2$^{\prime \prime}$-radius circular aperture \\
CNT4 & Number of \emph{Gaia} sources within a 4$^{\prime \prime}$-radius circular aperture \\
W1-W2 & unWISE $W$1-$W$2 colour \\
G-W1 & Extinction-corrected $G$-$W$1 colour \\
GW\_SEP & Separation (in arcsec) between a \emph{Gaia} source and its unWISE counterpart \\
\hline
\hline
\end{tabular}
\end{table*}

315 of the 718 extragalactic sources in the Bo\"{o}tes sub-region satisfy the requirement of $(F > 10$ OR $\hat{a} > 0.5)$ and are considered as AGNs, which implies that the AGN number density in the \emph{Gaia}-unWISE sample is $\sim$100 deg$^{-2}$ in the Bo\"{o}tes field. Considering that the Bo\"{o}tes field is among the deepest and most complete regions in the current \emph{Gaia}-unWISE sample with $G_{\rm peak} \approx 20.1$ mag and $G_{99} \approx 21.2$ mag, the overall AGN number density in the \emph{Gaia}-unWISE sample is expected to be less than $\sim$100 deg$^{-2}$. It also suggests that $\approx$99.5\% of the 641 million \emph{Gaia}-unWISE sources will be non-AGNs. An efficient and clean way of selecting AGNs from the \emph{Gaia}-unWISE sample is thus highly necessary. 

\section{Methodology}
\label{sect:method}
\subsection{Random Forest Algorithm}

In this work, we use the random forest (RF) algorithm for AGN/non-AGN classification and AGN photometric redshift estimation. The RF is a widely used, supervised machine learning algorithm that has been shown to generate robust models and work efficiently with large data sets. 

The RF algorithm relies on an ensemble of decision trees to make predictions for both classification and regression problems \citep{Breiman2001}. The decision trees are built independently based on features (i.e. source properties in our case) of input data sets, which are different bootstrap samples of the original training set. The decision tree is grown in a top-down fashion. At each node of a decision tree, the data set is split into two subsets according to the feature among a randomly-selected subset of all features that gives the highest information gain. The nodes are grown recursively until the stopping criterion is met. In a classification problem, each tree will calculate the probability (1 or 0) of an input object belonging to a particular class, and the mean class probability of all the trees is returned. In a regression problem, each tree will make a prediction on the unknown quantity that we are interested (photometric redshift in our case), and the average value from all the trees is used as the final estimation. 

The RF algorithm has been successfully applied to a variety of tasks in astronomy \citep[e.g.,][]{Carliles10, Dubath11, Richards12, Carrasco13, Wyrzykowski14, Jayasinghe19, Chen19}, including AGN classification and photometric redshift estimation \citep[e.g.,][]{Pichara12, Carrasco15, Schindler17, Nakoneczny19, Jin19}. We note that our work is the first RF-assisted AGN classification across the whole sky.  

\subsection{AGN Classification}

\subsubsection{Training and Test Sets}

We use \texttt{RandomForestClassifier} provided in the \texttt{scikit-learn} package \citep{scikit-learn} for AGN classification. We build the AGN data set for the RF classifier from the largest spectroscopically confirmed quasar sample --- the SDSS DR14 QSO catalogue \citep[DR14Q,][]{Paris18}. We perform a nearest neighbour cross-match between \emph{Gaia} DR2 and DR14Q using a matching radius of 0\farcs5, and find that 354,586 of the 526,356 quasars in DR14Q are detected and catalogued in \emph{Gaia} DR2. The unmatched DR14Q quasars are mostly fainter than $i \sim 20.2$ mag, beyond which \emph{Gaia} is significantly incomplete. Requiring unWISE counterparts within 2 arcsecs with non-zero fluxes in $W$1 results in 348,252 quasars, of which 339,194 (i.e. 97.4\%) further have non-zero fluxes in $W$2. We notice that some of the matched DR14Q quasars appear to have significant \emph{Gaia} parallaxes or proper motions, inconsistent with the fact that they should be stationary. After visual inspections of the images and spectra, we find that the majority of those ``moving'' quasars have close companions mostly due to projection effects, which affect the estimation of their parallaxes and proper motions. Consequently, parallax, proper motion, and photometry of those objects are no longer reliable, and may confuse the RF classifier. We therefore remove the 220 DR14Q quasars that have parallax or proper motion significance larger than 5$\sigma$. The remaining 338,974 quasars comprise the AGN data set and are also referred to as the \emph{Gaia}-unWISE-DR14 QSO sample.

To build the non-AGN data set, we randomly select 10 million objects from the \emph{Gaia}-Pan-STARRS1 crossmatch table and cross-match them with the unWISE catalogue using a matching radius of 2 arcsecs, which results in 2,351,443 objects with unWISE counterparts with non-zero $W$1 and $W$2 fluxes.
Obviously, we need to further clean this non-AGN data set by identifying and removing as many AGNs as possible. We therefore put together a known AGN compilation including almost 29 million known AGNs and AGN candidates (duplicates not removed) from the million quasar catalogue, version 5.7 \citep[MILLIQUAS,][]{Flesch15}, the AllWISE two-colour selected AGN catalogue \citep{Secrest15}, and the AllWISE R90 and C75 AGN catalogues \citep{Assef18}. We then remove the 10,902 objects in the non-AGN data set that have counterparts in the known AGN compilation within an aggressive matching radius of 5 arcsecs and are therefore potential AGNs. This number is consistent with the expectation based on the AGN/non-AGN fraction found in Section~\ref{sect:number_density}, which suggests $\sim$11500 AGNs in this data set. The cleaned non-AGN data set now has 2,340,541 objects.

The AGN data set and the cleaned non-AGN data set together make up the full data set for the RF classifier, which includes 2,679,515 objects. The full data set is shuffled and randomly split so that 80\% is used as the training set and the remaining 20\% is used as the test set. The training set contains 271,218 AGNs and 1,872,394 non-AGNs, while the test set contains 67,756 AGNs and 468,147 non-AGNs. 

\subsubsection{Feature Selection}

The RF classifier relies on a set of features (i.e. source properties) to determine whether a source is an AGN or not. In this work, we consider 16 features that we think are relevant in separating AGNs from stars and compact star-forming regions in galaxies. The features are summarised and explained in Table~\ref{tb:features}. Most of the features are directly available from the \emph{Gaia} DR2 catalogue and the unWISE catalogue, and more detailed descriptions can be found in \citet{Lindegren18}, the \emph{Gaia} DR2 datamodel\footnote{\url{https://gea.esac.esa.int/archive/documentation/GDR2/Gaia_archive/chap_datamodel/sec_dm_main_tables/ssec_dm_gaia_source.html}}, and \citet{Schlafly19}. We apply extinction corrections to the \emph{Gaia} $G$, BP, and RP magnitudes according to the extinction laws in \citet{Cardelli89} and \citet{O'Donnell94}, with the E($B-V$) value along each sight line extracted from the extinction map in \citet{Schlegel98}. \emph{Gaia} DR2 does not report parallax or proper motion for some sources (see \citet{Lindegren18} for details), and BP or RP under certain circumstances (see \citet{Riello18} for details). We flag those null parallaxes and proper motions as $-$999, and null BP$-G$, $G-$RP, or BP$-$RP colours as 999. In the full data set, 61,332 AGNs and 198,208 non-AGNs do no have parallaxes and proper motions, 21,273 AGNs and 197,300 non-AGNs do not have BP-$G$ colours, 21,252 AGNs and 196,511 non-AGNs do not have $G-$RP colours, and 21,285 AGNs and 197,637 non-AGNs do not have BP$-$RP colours. 

Following \citet{Belokurov17}, we construct one feature, G\_VAR, from direct measurements as
\begin{equation}
    {\text{G\_VAR}} = \sqrt{{\tt PHOT\_G\_N\_OBS}} \times \frac{{\tt PHOT\_G\_MEAN\_FLUX\_ERROR}}{{\tt PHOT\_G\_MEAN\_FLUX}},
\end{equation}
in which {\tt PHOT\_G\_N\_OBS} is the number of observations contributing to $G$ photometry, {\tt PHOT\_G\_MEAN\_FLUX} is the $G$-band mean flux, and {\tt PHOT\_G\_MEAN\_FLUX\_ERROR} is the standard deviation of the $G$-band flux divided by $\sqrt{{\tt PHOT\_G\_N\_OBS}}$. Clearly, $2.5 \times$G\_VAR/$\ln(10)$ is equivalent to the variation in the $G$-band magnitude. It is therefore helpful to include this feature, which should encode a source's variability information during the observing epochs. 
However, some other technical effects can also lead to a substantial variation in the $G$-band flux, for instance, a mix of different \emph{Gaia} scanning directions, especially for extended sources with non-circular surface brightness distributions. 
For each source, we compute the numbers of \emph{Gaia} sources (the target source is included) within circular apertures of 1$^{\prime \prime}$, 2$^{\prime \prime}$, and 4$^{\prime \prime}$ radii and denote them as CNT1, CNT2, and CNT4, respectively. These three features, together with the separation between a \emph{Gaia} source and its unWISE counterpart, GW\_SEP, provide a measure of the local crowding effect and the robustness of the \emph{Gaia} astrometric solution and \emph{Gaia} and unWISE photometric measurements, and thus help in better classifying a source. 

To select the most important/relevant features for AGN classification, we first train a \texttt{RandomForestClassifier} with its default parameter choices with the training set using all the features listed in Table~\ref{tb:features}, and record its performance on the test set as measured by the \texttt{f1\_score} metric. The F1 score is defined as 
\begin{equation}
    \text{F1} = 2 \times \frac{\text{completeness} \times \text{reliability}}{\text{completeness}+\text{reliability}}.
\end{equation}
For example, suppose a data set contains 100 AGNs and 10000 non-AGNs. For a classifier that mis-classifies 1 AGN and 10 non-AGNs, the F1 score is 0.947. 
We choose to optimise the classifier for the F1 score because it measures both the completeness and reliability. For this baseline model using 16 features, the F1 score is 0.9875. 
The relative importance of the 16 features is returned by the \texttt{feature\_importances\_} attribute of the \texttt{RandomForestClassifier} method. We remove 4 features (i.e., AEN, GOF, CNT2, and CNT1) that have a cumulative importance less than 0.01, and re-train the model. The F1 score of the new model is 0.9874, i.e. nearly as good as the baseline model. We therefore only use the remaining 12 features for the AGN classification. 

\subsubsection{Classifier Tuning \& Performance}

RF classifiers require specification of a number of parameters describing what kinds of trees may be built. Fortunately, we find that we can obtain clean samples of AGNs over a wide range of RF parameters. Nevertheless, we select the best possible RF parameters by optimising the RF performance over the four parameters, i.e. \emph{max\_features}, \emph{max\_depth}, \emph{class\_weight}, and \emph{criterion}, that are most relevant to the classifier's performance in our case. We refer interested readers to the \texttt{scikit-learn} documentation\footnote{\url{https://scikit-learn.org/stable/documentation.html}} for a full description of the role of the parameters. We consider \emph{max\_features} = [ 3, 4, 5, 6 ], \emph{max\_depth} = [ None, 25, 50 ], \emph{class\_weight} = [ None, balanced, \{0:1, 1:100\}, \{0:1, 1:200\}, \{0:1, 1:500\}, \{0:1, 1:1000\}, \{0:1, 1:10000\} ], and \emph{criterion} = [ entropy, gini ]. The remaining parameters of \texttt{RandomForestClassifier} are set to their default values. We find that the combination of parameters that gives the highest F1 score is \emph{max\_features} = 3, \emph{max\_depth} = 50, \emph{class\_weight} = \{0:1, 1:200\}, and \emph{criterion} =  entropy. We therefore adopt these choices and obtain the best-trained AGN classifier after training on the training set. Nevertheless, we note that changes in the F1 score for the considered various parameter combinations are very tiny, on the level of 0.001. 

\begin{figure*}
    \centering
    \includegraphics[width=0.98\textwidth]{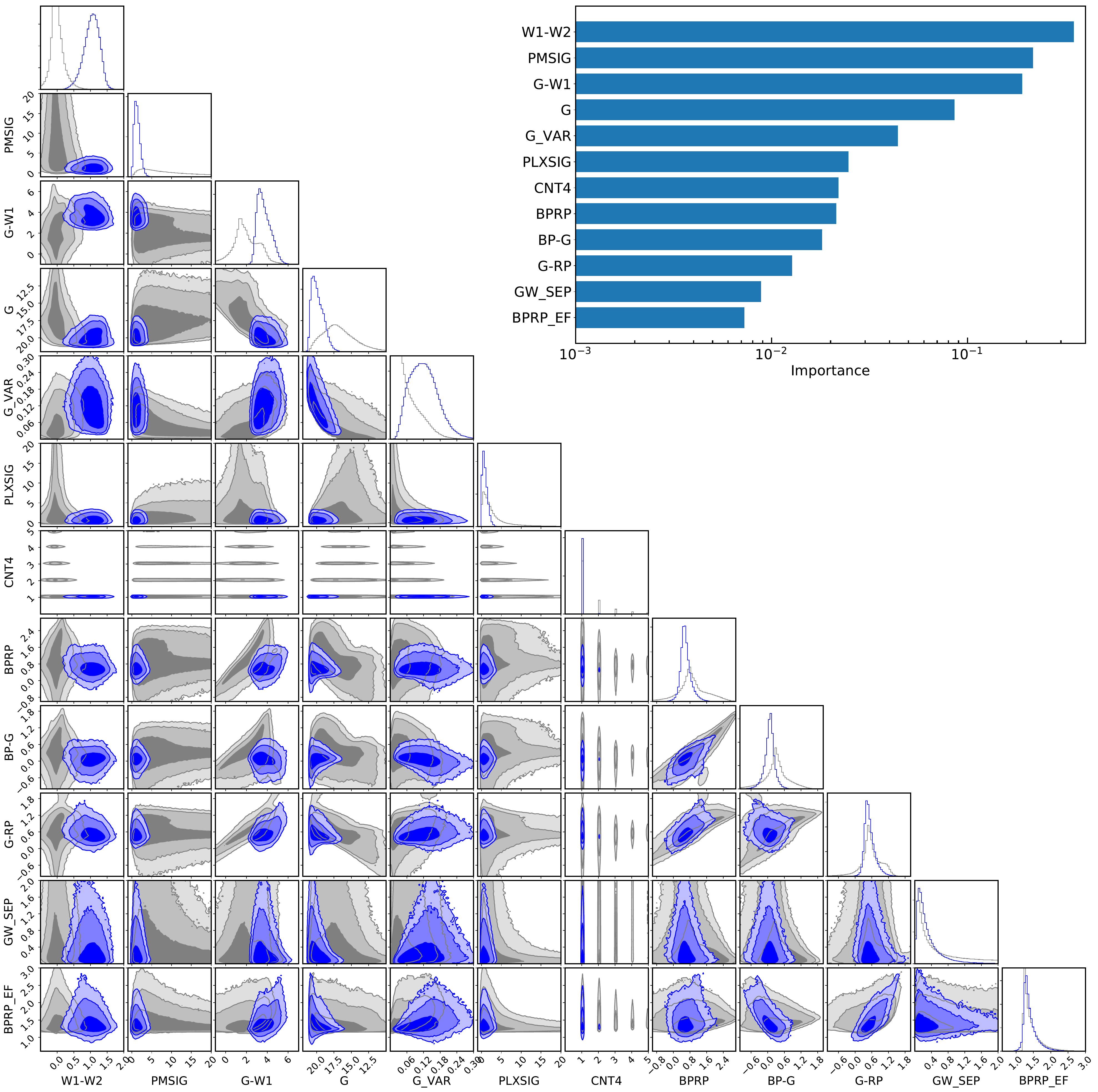}
    \caption{\label{fig:features} \emph{Upper corner:} Relative importance of the 12 features used by the best-trained AGN classifier. \emph{Lower corner:} Two-dimensional distributions and one-dimensional histograms of AGNs (blue) and non-AGNs (grey) in the training set in various feature spaces. The contours enclose 68\%, 95\%, and 99\% of AGNs and non-AGNs. The features are ordered by the relative importance. }
\end{figure*}

The relative importance, in descending order, of the 12 features used in the best-trained AGN classifier is shown in the upper corner of Figure~\ref{fig:features}. 
The lower corner of Figure~\ref{fig:features} shows the two-dimensional distributions and one-dimensional histograms of the 12 features, ordered by the importance, for AGNs and non-AGNs in the training set. The first thing to notice is the clear separation between AGNs and non-AGNs in the $W$1-$W$2 colour, which confirms again the effectiveness of $W$1-$W$2 colour in distinguishing AGNs from stars and galaxies. The PMSIG distribution is also different for AGNs and non-AGNs, with non-AGNs having an extended tail towards large PMSIG due to the presence of moving stars. The $G-W1$ colour of AGNs in the training set peaks around 4, with more than 95\% having (extinction-uncorrected) $G-W1 > 2.9$. The $G-W1$ colour of non-AGNs show a bimodal distribution, with the bluer component contributed mostly by stars and the redder component mostly by galaxies. Recent work by \citet{Lemon19} showed that one can efficiently distinguish QSOs and strongly-lensed QSOs from stars using the combination of $W1-W2$ and $G-W1$ colours. As expected, AGNs generally have larger G\_VAR with a peak value of $\approx 0.12$, or 0.13 mag, while non-AGNs peaks at G\_VAR $\approx 0.01$. 

\begin{table}
\centering
\caption{\label{tb:performance} Performance of our AGN classifier based on the test set. The definitions of the true positive rate or TPR and false positive rate or FPR are explained in the text. A good classifier should deliver a high TPR and maintain a low FPR at the same time. For comparison, we also show the results of applying the \emph{WISE}-only AGN selection criteria used in \citet{Stern12} and \citet{Assef18} to the same test set. Our AGN classifier delivers significantly better performance.}
\begin{tabular}{l c c}
\hline
\hline
& TPR & FPR \\
\hline
This work, $P_{\rm RF} \geq 0.69$ & 98.10\% & 0.15\% \\
This work, $P_{\rm RF} \geq 0.94$ & 92.73\% & 0.08\% \\
\hline
\citet{Stern12} & 84.03\% & 0.34\% \\
\citet{Assef18}, C75 & 90.63\% & 0.58\% \\
\citet{Assef18}, R90 & 60.67\% & 0.17\% \\
\hline
\hline
\end{tabular}
\end{table}

Table~\ref{tb:performance} presents the performance of the best-trained AGN classifier when applied to the test set. The true positive rate (TPR, equivalent to completeness) is the fraction of AGNs that are classified as AGNs, while the false positive rate (FPR) is the fraction of non-AGNs that are mis-classified as AGNs. A good classifier should deliver a high TPR and maintain a low FPR at the same time. We show two sets of results that correspond to two different AGN probability thresholds, which, as will be shown later, yield 75\% completeness ($P_{\rm RF} \geq 0.69$) and 85\% reliability ($P_{\rm RF} \geq 0.94$) respectively. For the test set, the best-trained AGN classifier achieves a TPR of $\gtrsim 93\%$, and the FPR is 0.08\%--0.15\%. 

To illustrate the advantage of combining \emph{Gaia} (optical) and \emph{WISE} (mid-IR) data in identifying AGNs, we apply the \emph{WISE}-only AGN selection criteria used in \citet{Stern12} and \citet{Assef18} to the same test set. More specifically, sources are classified as AGNs if they satisfy $W1-W2 \geq 0.8$ (\citet{Stern12}), or $W1-W2 > 0.71$ (the C75 criterion used by \citet{Assef18} to achieve 75\% completeness), or 
\[
    W1-W2 > 
\begin{cases}
    0.650 \times e^{[0.153 \times (W2-13.86)^2]},& W2 > 13.86 \\
    0.650,              & W2 \leq 13.86
\end{cases}
\] (the R90 criterion used by \citet{Assef18} to achieve 90\% reliability). It is clear that using optical and mid-IR data, the TPR becomes significantly higher, i.e. more AGNs can be identified. More importantly, our FPRs are lower by 0.25\% on average than those of the \emph{WISE}-only criteria (the R90 criterion in \citet{Assef18} achieves a comparably small FPR, but at the cost of a substantially lower TPR). Although the improvement of $\sim$0.25\% in the FPR seems tiny, it will lead to a huge improvement in the reliability because the number of non-AGNs in the \emph{Gaia}-unWISE sample is almost 640 million. If assuming the non-AGN test set is representative of the non-AGNs in the \emph{Gaia}-unWISE sample, an improvement of 0.25\% in the FPR can prevent $\approx$1.6 million non-AGNs being mis-classified as AGNs.

\subsection{Photometric Redshift Estimation}

\begin{figure}
    \centering
    \includegraphics[width=0.45\textwidth]{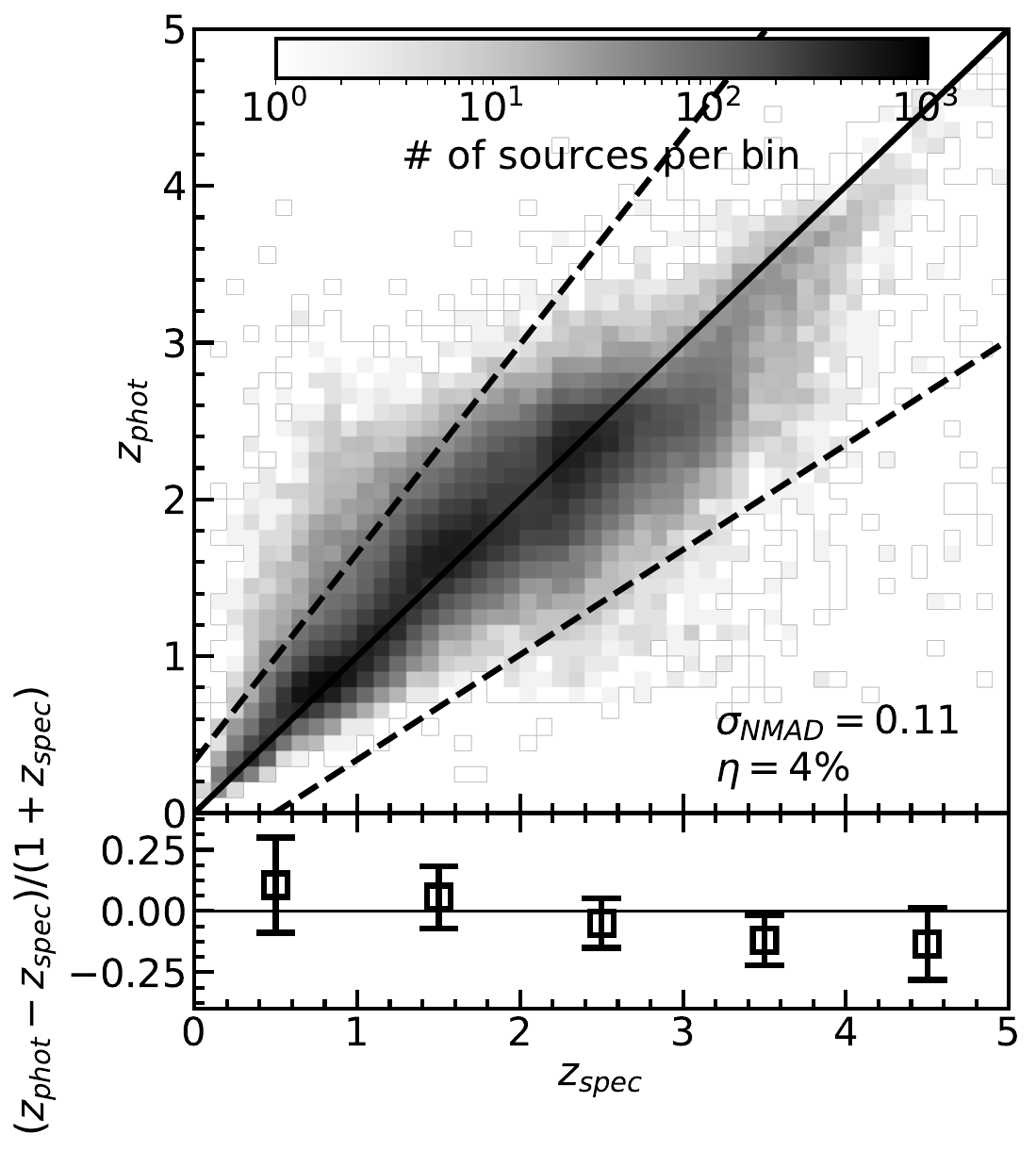}
    \caption{\emph{Top}: Comparison between the estimated photometric redshift and the spectroscopic redshift for the test set. The overall two-dimensional histogram follows the solid one-to-one line, and the photometric redshift accuracy $\sigma_{\rm NMAD}$ is 0.11. 4\% of the objects fall outside the region bounded by the two dashed lines, and are referred to as catastrophic outliers. \emph{Bottom}: The mean and 1$\sigma$ dispersion of the fractional difference $(z_{\rm phot}-z_{\rm spec})/(1+z_{\rm spec})$ in five redshift bins. A mild bias at $\lesssim 1\sigma$ level is seen, suggesting the photometric redshifts tend to be over-estimated for low-redshift AGNs and under-estimated for high-redshift AGNs.}
    \label{fig:phot_z_performance}
\end{figure}

We use \texttt{RandomForestRegressor} provided in the \texttt{scikit-learn} package for the photometric redshift estimation. 
80\% of the \emph{Gaia}-unWISE-DR14 QSO sample (271,179 AGNs) is randomly chosen as the training set, and the remaining 20\% is used as the test set. The 10 features that are used in the RF regressor are G, W1, BP-G, BP-RP, G-RP, G-W1, RP-W1, W1-W2, G\_VAR, and GW\_SEP. The RP-W1 feature is derived from G-W1 and G-RP. Again, we find that similar photometric-redshift accuracy can be achieved for a wide range of RF parameters. Nevertheless, we optimise the choices for the two parameters, i.e. \emph{max\_features} and \emph{max\_depth}, that are usually most relevant to a regressor's performance. In particular, we consider \emph{max\_features} = [ 2, 3, 4, 6, 8, 10 ] and \emph{max\_depth} = [ None, 10, 25, 50 ]. The remaining parameters of \texttt{RandomForestRegressor} are set to their default values. 

We use the standard $R^2$ score to evaluate the performance of the RF regressor. Assuming the true, spectroscopic redshifts are denoted as $z_{\rm spec}^i$, the mean of $z_{\rm spec}^i$ is denoted as $\bar{z}$, and the predicted redshifts are denoted as $z_{\rm phot}^i$, the $R^2$ score (also known as the coefficient of determination) is defined as 
\begin{equation}
    R^2 \equiv 1 - \frac{\sum_i (z_{\rm spec}^i - z_{\rm phot}^i)^2}{\sum_i (z_{\rm spec}^i - \bar{z})^2}. 
\end{equation}
Clearly, the best $R^2$ score is 1. The combination of parameters that gives the highest $R^2$ score of $0.752$ is \emph{max\_features} = 4 and \emph{max\_depth} = 25. Nevertheless, changes in the $R^2$ score for the considered parameter combinations are very tiny. For example, a RF regressor with all its parameters set to default values delivers a $R^2$ score of 0.749. In the best-trained RF regressor, the most important feature is RP-W1 (relative importance of 0.22), followed by G-W1, W1-W2, W1, BP-G, GW\_SEP, BP-RP, G, G\_VAR, and G-RP. 

Following the convention in the literature \citep[e.g.,][]{Ilbert09, Ananna17, Fotopoulou18}, we estimate the photometric redshift accuracy using the normalised median absolute deviation defined as 
\begin{equation}
    \sigma_{\rm NMAD} = 1.48 \times \text{median } (\frac{\left |z_{\rm phot}^i-z_{\rm spec}^i \right |}{1+z_{\rm spec}^i}). 
\end{equation}
The top panel in Figure~\ref{fig:phot_z_performance} shows the comparison between $z_{\rm phot}$ from the best-trained RF regressor and $z_{\rm spec}$ of the test set. The overall distribution is centred on the one-to-one relation (solid black line), and $\sigma_{\rm NMAD} = 0.11$. We estimate the rate of catastrophic outliers $\eta$ as the fraction of sources that have 
\begin{equation}
    \frac{\left |z_{\rm phot}-z_{\rm spec} \right |}{1+z_{\rm spec}} > 3 \times \sigma_{\rm NMAD} = 0.33.
\end{equation}
The two dashed lines indicate the boundary where ${\left |z_{\rm phot}-z_{\rm spec} \right |} > 0.33 \times (1+z_{\rm spec})$. The rate of catastrophic outliers is $\eta = 4\%$. We further divide the test set into five equally-spaced redshift bins from 0 to 5, and find that the mean and standard deviation of $(z_{\rm phot}-z_{\rm spec})/(1+z_{\rm spec})$ is $0.10 \pm 0.19$ for $0 < z_{\rm spec} \leq 1$, $0.06 \pm 0.13$ for $1 < z_{\rm spec} \leq 2$, $-0.05 \pm 0.10$ for $2 < z_{\rm spec} \leq 3$, $-0.12 \pm 0.10$ for $3 < z_{\rm spec} \leq 4$, $-0.13 \pm 0.15$ for $4 < z_{\rm spec} \leq 5$ (bottom panel in Figure~\ref{fig:phot_z_performance}), which implies a mild bias (at $\lesssim 1\sigma$ level) in the sense that our best-trained regressor tends to over-estimate the redshifts for AGNs at $z \lesssim 2$ and under-estimate the redshifts for AGNs at $z \gtrsim 3$. We have tried two other commonly used, machine-learning based regression methods, i.e. XGBoost \citep{Chen16} and Support Vector Regression. They deliver very similar photometric redshift accuracy as the RF regressor, and the bias persists. It suggests that this bias is due to the intrinsic uncertainties in the AGN photometric redshift estimation rather than the choices of the regression method or the parameter settings, especially when only broadband colours are used. 

Nevertheless, the photometric redshift accuracy is comparable to performances of recent work on AGN photometric redshift estimation, most of which use more colours than our photometric redshift estimator~\citep[e.g.,][]{Maddox12, Chung14, Schindler17, Jin19}. We thus use the best-trained RF regressor to estimate the photometric redshifts of our AGN candidates. 

\section{Results}
\label{sect:results}

We apply the best-trained AGN classifier to the \emph{Gaia}-unWISE sample of 641,266,363 sources and obtain 3,175,537 sources with AGN probability $P_{\rm RF} \geq 0.5$, which we refer to as AGN candidates. Upon visual inspections, we notice significant over-densities of AGN candidates towards the directions of the Large Magellanic Cloud (LMC) and Small Magellanic Cloud (SMC). Querying against the SIMBAD database finds that the majority of those AGN candidates are actually YSOs and AGB stars in the LMC and SMC that have AGN-like $W1-W2$ colours \citep[e.g.,][]{Nikutta14}. Because of the extremely high source densities in these nearby galaxies, the \emph{Gaia} and \emph{WISE} photometry become less reliable. We therefore remove AGN candidates that are located within twice the radius of LMC, SMC, and M31, which is the nearest big galaxy to the Milky Way. The central positions and radii of LMC, SMC, and M31 are taken from the Catalog and Atlas of the Local Volume Galaxies \citep{Karachentsev13}. This step removes an area of 541 deg$^2$. The total number of AGN candidates with $P_{\rm RF} \geq 0.5$ is reduced to 3,104,739, which is referred to as the raw AGN catalogue. 

In this work, we will construct two AGN catalogues out of the raw AGN catalogue that are optimised for completeness and reliability respectively. We now explain how this can be achieved by imposing simple cuts on $P_{\rm RF}$.

\subsection{C75 And R85 AGN Catalogues}

\begin{figure}
    \centering
    \includegraphics[width=0.48\textwidth]{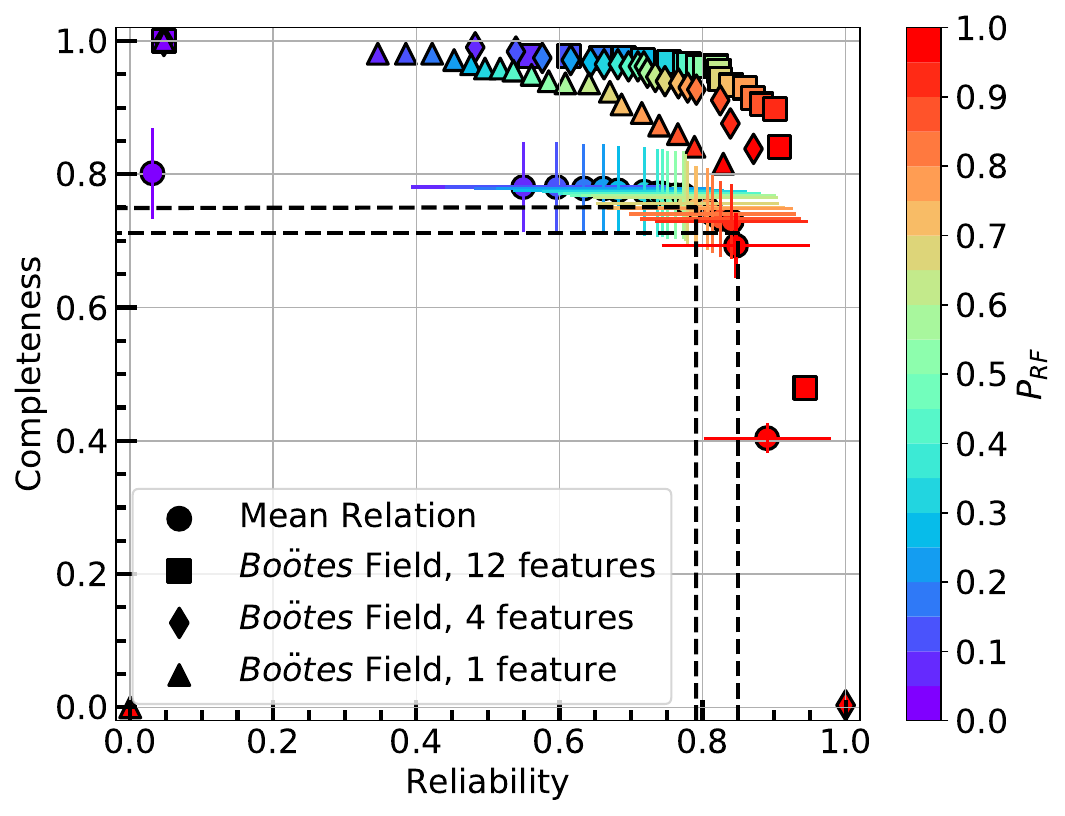}
    \caption{\label{fig:completeness_reliability} The mean completeness-reliability relation (filled circles) derived from 100 spatially randomly-distributed test fields. The error bars correspond to $1\sigma$ variations in completeness and reliability. Squares represent the completeness-reliability relation in the Bo\"{o}tes reference field obtained by the best-trained RF classifier. Diamonds and triangles represent the same relation obtained by two other RF classifiers using fewer features. The symbols are colour-coded according to the $P_{\rm RF}$ threshold. The dashed lines highlight two $P_{\rm RF}$ thresholds at which the mean completeness reaches 75\% ($P_{\rm RF} \geq 0.69$) and the mean reliability reaches 85\% ($P_{\rm RF} \geq 0.94$). }
\end{figure}

We use the Bo\"{o}tes field, which is among the deepest fields in the \emph{Gaia}-unWISE sample, as a reference to estimate the overall completeness and reliability of the final AGN catalogue at different $P_{\rm RF}$ thresholds. We construct a reference sample including all the 6,703 sources in the \emph{Gaia}-unWISE sample that fall within the previously-defined $2^{\circ} \times 2^{\circ}$ sub-region in the Bo\"{o}tes field (denoted as the reference field). For every source in the reference sample, we obtain its AGN probability $P_{\rm RF}$ from the best-trained AGN classifier. On the other hand, a nearest neighbour cross-match using a matching radius of 0\farcs6 finds that 4,523 sources in the reference sample are also in the Bo\"{o}tes source catalogue, for which we can decide whether they are AGNs based on the $F$ ratio and $\hat{a}$ parameter requirement. The unmatched ones are mostly bright objects that were not included in the Bo\"{o}tes source catalogue due to the saturation limit/incompleteness, which we conservatively assume to be non-AGNs. At any given $P_{\rm RF}$ threshold, we can compute the number of sources in the reference sample that have $P_{\rm RF}$ larger than or equal to the threshold (denoted as $N_{\rm candidate}$) and the number of sources among those candidates that satisfy the $(F > 10$ OR $\hat{a} > 0.5)$ criterion (denoted as $N_{\rm AGN}$). In addition, we know from Section~\ref{sect:number_density} that the total number of AGNs in this reference field is 315. The completeness is therefore given by $N_{\rm AGN}/315$, and the reliability is given by $N_{\rm AGN}/N_{\rm candidate}$. The squares in Figure~\ref{fig:completeness_reliability} correspond to the completeness-reliability relation at different $P_{\rm RF}$ thresholds in the Bo\"{o}tes reference field. We note that the actual reliability should be slightly higher than the inferred values because of the adopted conservative treatment of the unmatched objects in the reference sample. 

Clearly, the completeness-reliability relation derived from the deep Bo\"{o}tes reference field will be optimistic for the final AGN catalogue. Nevertheless, due to the lack of Bo\"{o}tes-like fields with sufficient and representative sky coverage, we choose to estimate the overall completeness and reliability of the final AGN catalogue through simulations. In particular, we select 100 test fields with the same area as the reference field that are randomly distributed across the high-latitude sky (i.e. $\left | b \right | > 20^{\circ}$). We adopt this requirement because the majority of the raw AGN catalogue are distributed at $\left | b \right | > 20^{\circ}$. 
For each test field, we first obtain the \emph{Gaia} $G$-band magnitude distribution, d$N$/d$G$ (test), for all the \emph{Gaia}-unWISE sources therein. A mock sample is generated by resampling the reference sample to match d$N$/d$G$ (test). Because the source density in the test field can be different from that of the reference field, we adjust the relative weight of non-AGNs to AGNs in the reference sample to $w_{\rm non-AGN} \equiv [N_{\rm source} \text{(test)} - 315] / [N_{\rm source} \text{(ref)} - 315]$, in which $N_{\rm source} \text{(test)}$ and $N_{\rm source} \text{(ref)}$ are the total number of \emph{Gaia}-unWISE sources in the test field and reference field and 315 is the total number of AGNs in the reference field. As a result, the probability of selecting a non-AGN from the reference sample is a factor of $w_{\rm non-AGN}$ larger than the probability of selecting an AGN in the resampling process. 
For each test field, 100 independent mock samples are generated. We compute the completeness-reliability relation for each mock sample following the above procedures for the Bo\"{o}tes reference field, and take the mean completeness-reliability relation as the relation for this test field. This process is done for all the 100 test fields. 

\begin{figure}
    \centering
    \includegraphics[width=0.48\textwidth]{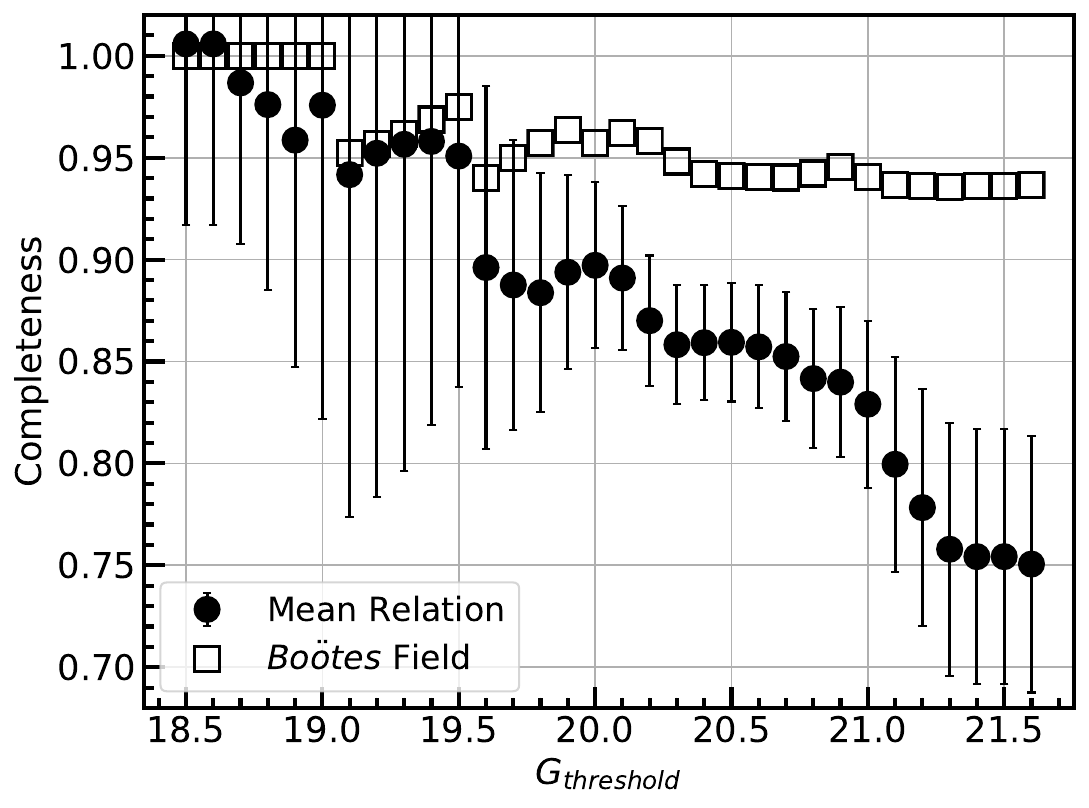}
    \caption{\label{fig:completeness_Glimit} Mean completeness at $P_{\rm RF} \geq 0.69$ of the 100 test fields (filled circles) and the completeness of the Bo\"{o}tes reference field (squares) as a function of the $G$ magnitude threshold. }
\end{figure}

The circles in Figure~\ref{fig:completeness_reliability} show the mean completeness-reliability relation for the 100 test fields, and the error bars represent the $1\sigma$ standard deviations in completeness and reliability. The colour of the circles corresponds to the $P_{\rm RF}$ threshold. The completeness and reliability vary significantly across the test fields, on the levels of $\sim$7\% and $\sim$13\% respectively, due to the spatial variations of source density and \emph{Gaia}-unWISE completeness and limiting magnitude. 
We find that the mean completeness reaches at least 75\% (mean reliability $\sim 79\%$) at $P_{\rm RF} \geq 0.69$, and the mean reliability reaches at least 85\% (mean completeness $\sim 71\%$) at $P_{\rm RF} \geq 0.94$. We therefore construct two AGN catalogues, denoted as C75 and R85, by selecting AGN candidates of $P_{\rm RF} \geq 0.69$ and $P_{\rm RF} \geq 0.94$ respectively. The C75 AGN catalogue contains 2,734,464 sources, and the R85 AGN catalogue contains 2,182,193 sources. It is obvious that the R85 catalogue is a subset of the C75 catalogue. The C75 AGN catalogue is publicly available as a FITS file at \url{https://www.ast.cam.ac.uk/~ypshu/AGN_Catalogues.html}. 

\begin{figure*}
    \centering
    \includegraphics[width=0.8\textwidth]{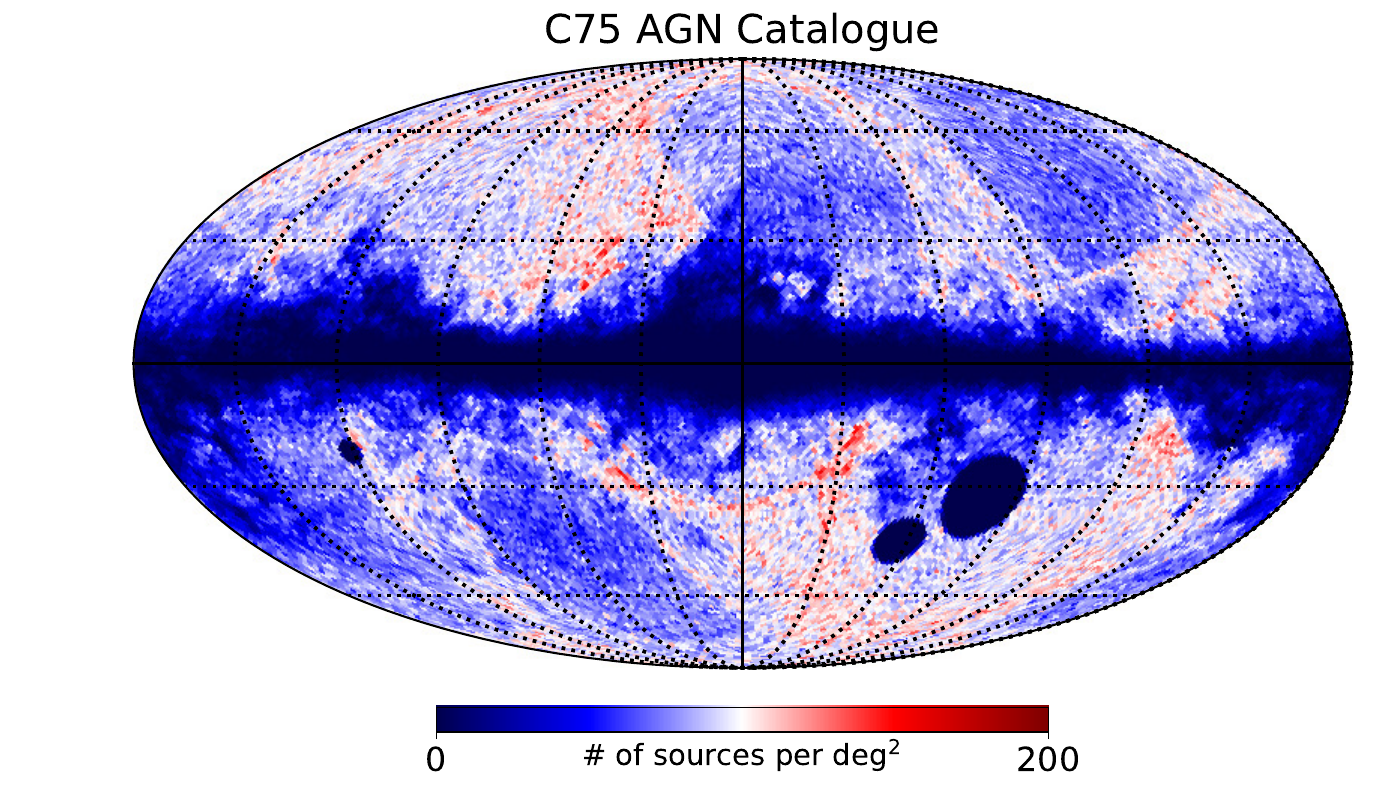}
    \includegraphics[width=0.8\textwidth]{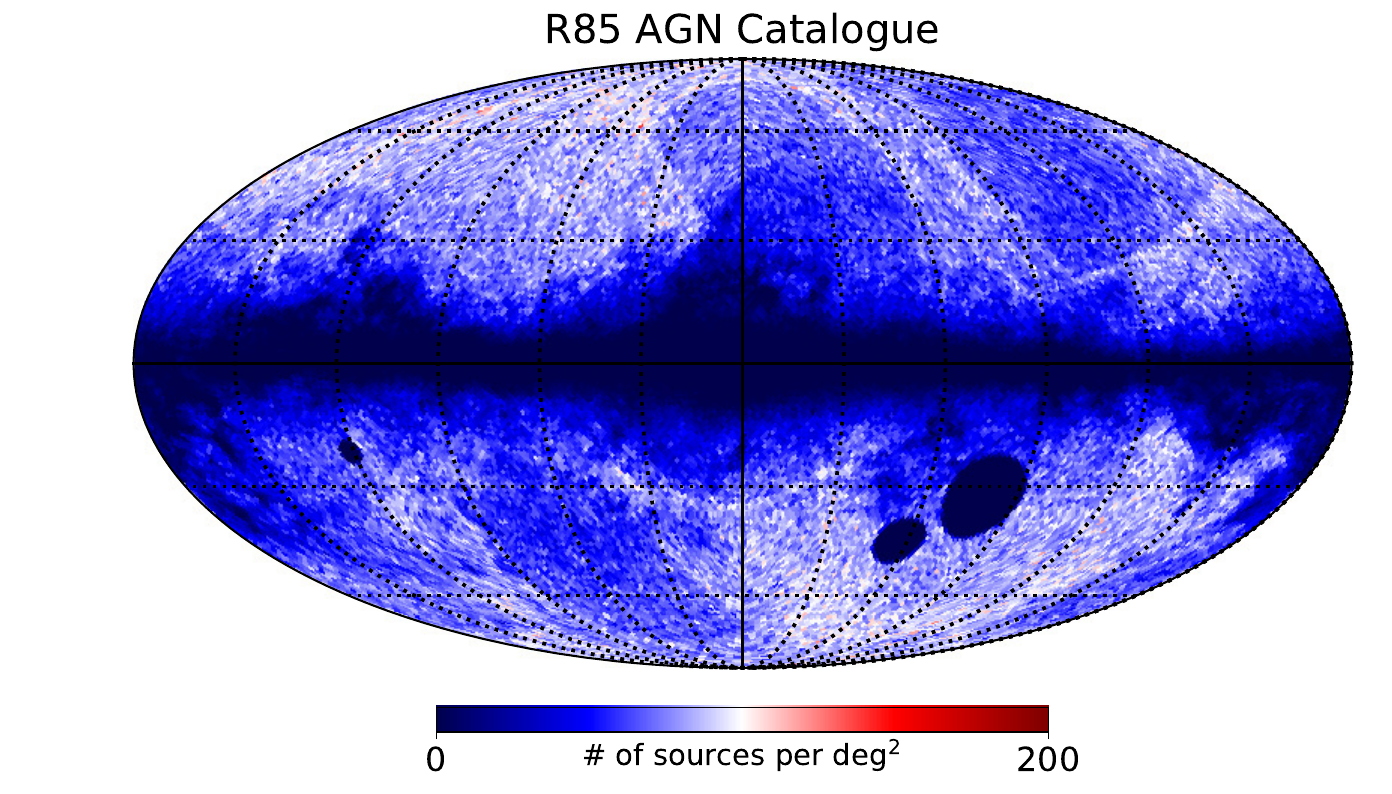}
    \caption{\label{fig:spatial_distribution} Spatial distributions (in Mollweide projection) of AGN candidates in the C75 (\emph{top}) and R85 (\emph{bottom}) AGN catalogues in the Galactic coordinate system.}
\end{figure*}

The completeness of our AGN catalogues is sensitive to the $G$ magnitude threshold. We can see from Figure~\ref{fig:completeness_Glimit} that although the estimated overall completeness is 75\%, the C75 catalogue is $\approx 95\%$ complete for AGN candidates at $G \leq 19.5$ mag and $\approx 90\%$ complete for AGN candidates at $G \leq 20$ mag. For the Bo\"{o}tes field that is among the deepest regions in the current \emph{Gaia}-unWISE sample, the completeness at $P_{\rm RF} \geq 0.69$ is about 93--100\%, and it varies very little with the $G$ magnitude threshold. We thus expect the overall completeness of AGN catalogues built from later \emph{Gaia} data releases to improve substantially to that of the Bo\"{o}tes field as more repeated \emph{Gaia} observations across the whole sky will be conducted.

To assess by how much the performance of the RF classifier degrades when fewer features are used, we consider two other RF classifiers that are trained on the top four most important features W1$-$W2, PMSIG, G$-$W1, and G and on the most important feature W1$-$W2 alone. The diamond and triangle symbols in Figure~\ref{fig:completeness_reliability} show the completeness-reliability relations in the Bo\"{o}tes reference field using $P_{\rm RF}$ values given by these two other RF classifiers respectively. RF classifiers trained on fewer features generally deliver lower completeness and reliability values. At $P_{\rm RF} \geq 0.69$, the RF classifier using 4 features has the same completeness of 93.6\% as the best-trained RF classifier using 12 features, while the RF classifier using only 1 feature has a lower completeness of 90.8\%. At $P_{\rm RF} \geq 0.94$, the best-trained RF classifier achieves a reliability of 90.7\%, while the other two RF classifiers deliver lower reliability of 86.0\% and 82.4\% respectively.

\subsection{Demographics of the AGN Candidates}

\begin{figure*}
    \centering
    \includegraphics[width=0.48\textwidth]{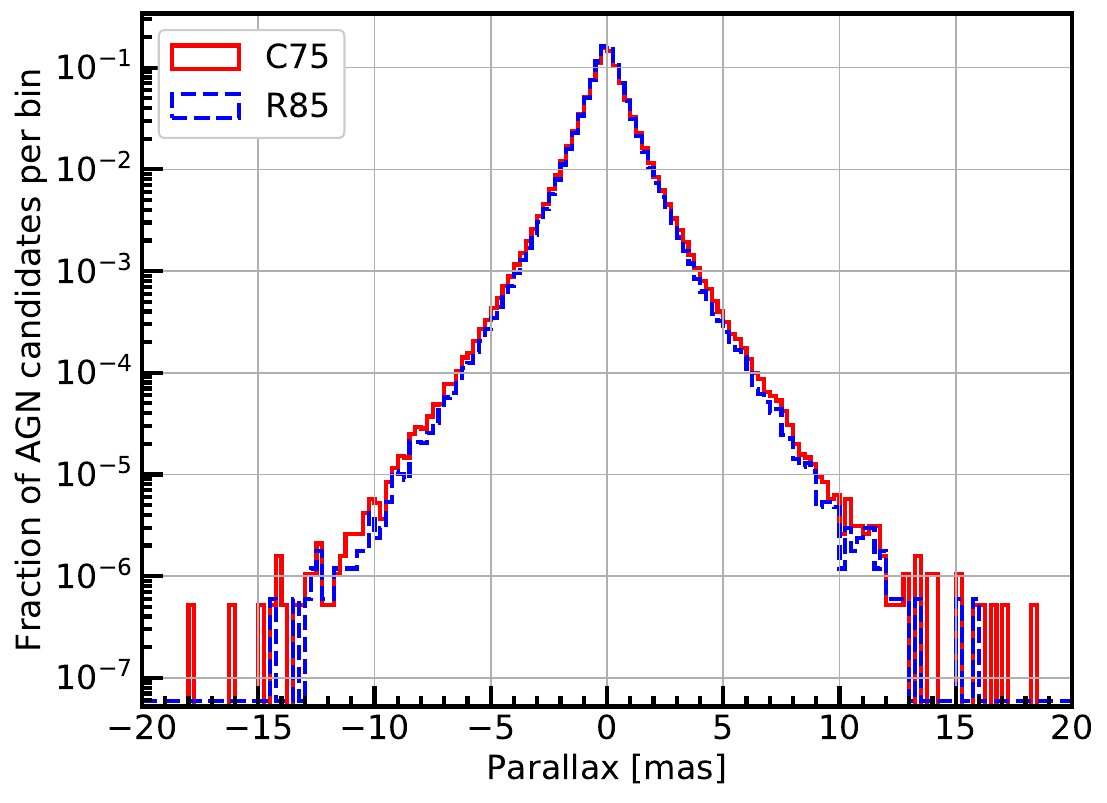}
    \includegraphics[width=0.48\textwidth]{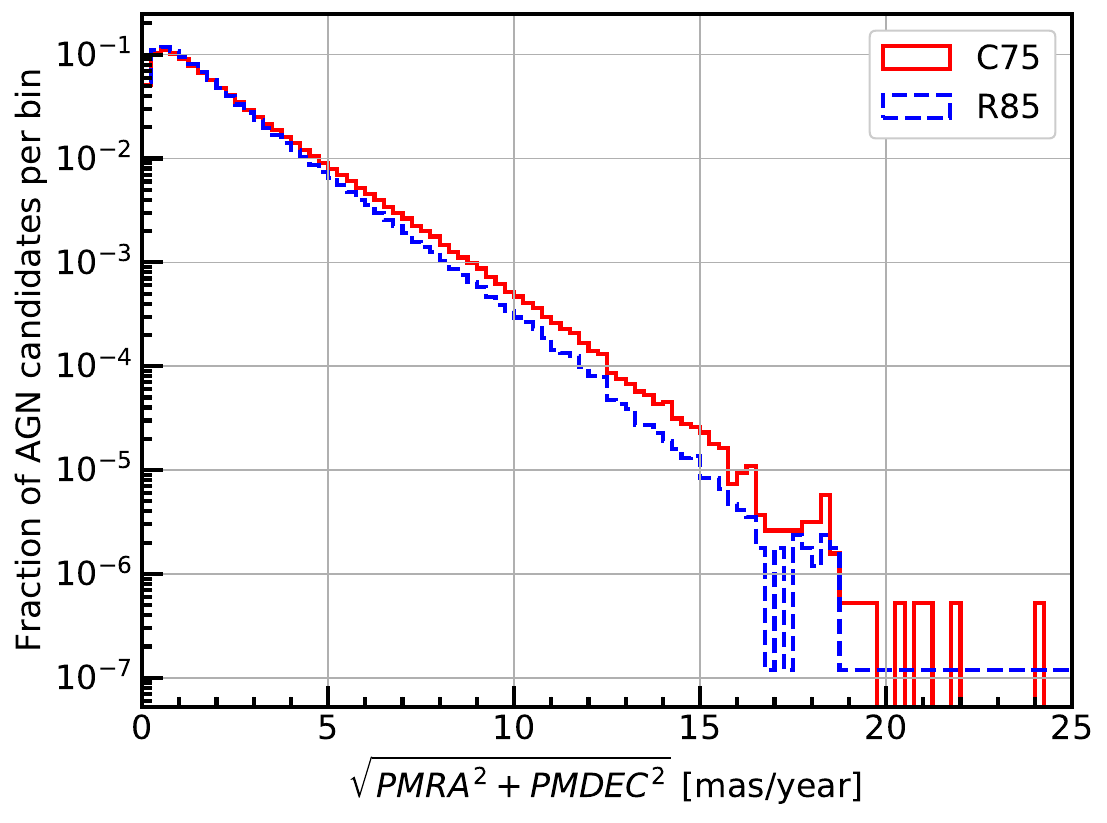}
    \includegraphics[width=0.48\textwidth]{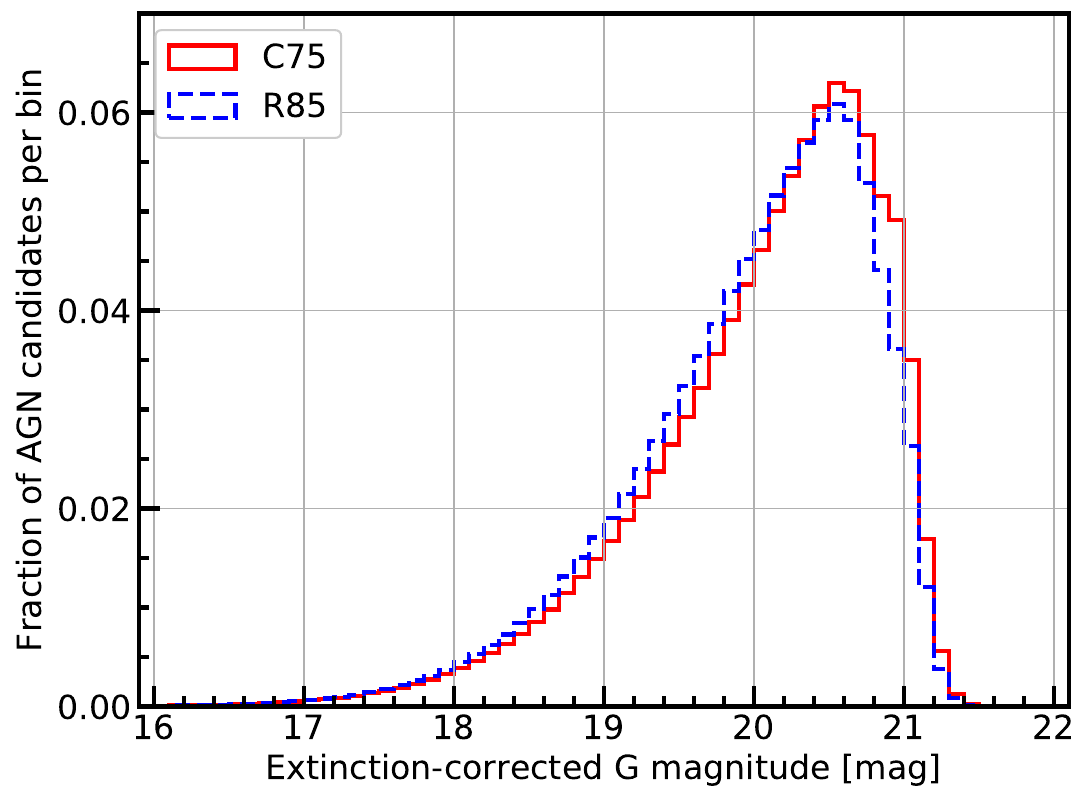}
    \includegraphics[width=0.48\textwidth]{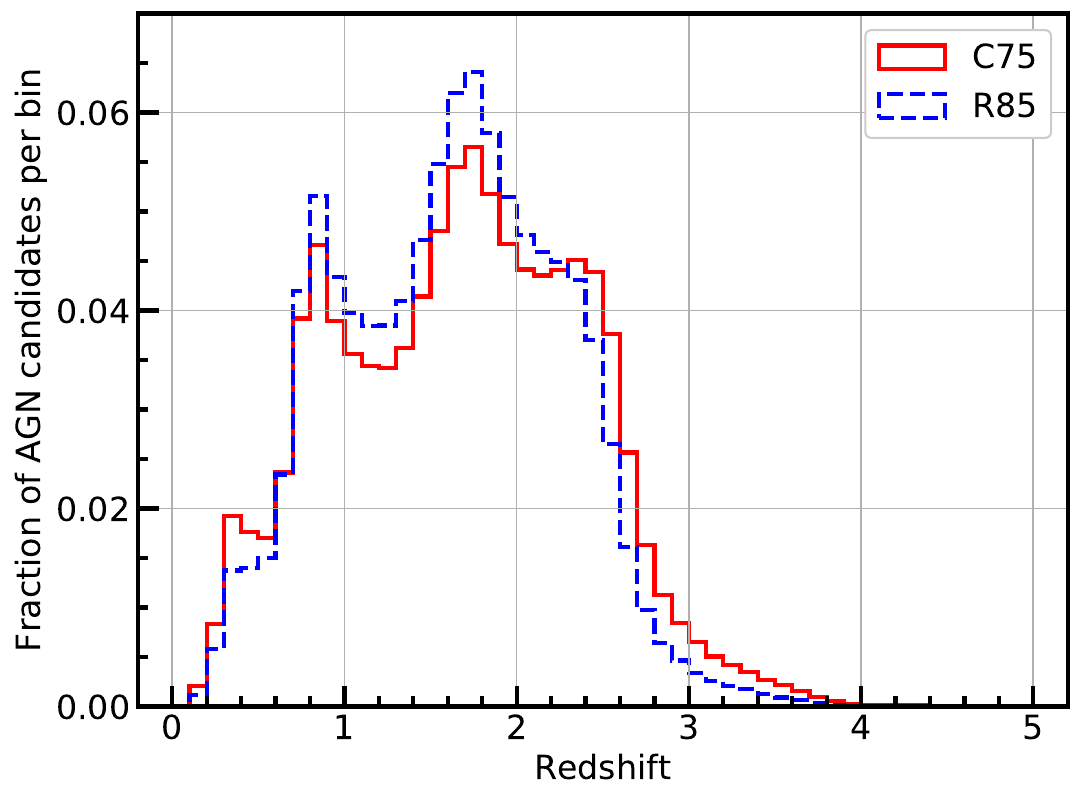}
    \caption{\label{fig:distributions} Normalised histograms of parallax (\emph{top left}), proper motion (\emph{top right}), extinction-corrected $G$-band magnitude (\emph{bottom left}), and photometric redshift (\emph{bottom right}) for the C75 (solid red) and R85 (dashed blue) AGN catalogues.}
\end{figure*}

Figure~\ref{fig:spatial_distribution} shows the spatial density distributions of the C75 and R85 AGN catalogues in the Galactic coordinate system. The colour scale is chosen such that white corresponds to an AGN density of 100 deg$^{-2}$ as estimated from the Bo\"{o}tes field, and redder or bluer colour corresponds to higher or lower densities. The first thing to notice is that the AGN density distributions of the C75 and R85 catalogues strongly correlate with the Galactic extinction distribution, and the AGN densities drop quickly to zero towards the Galactic plane and the bulge region, primarily because the high extinction in those regions prevents faint AGNs being detected when optical data are involved. In addition, this could be partially related to a selection bias in our model. The mean E($B-V$) value of the AGN training set is about 0.03 mag and more than 99\% AGNs in the training set have E($B-V$) $\leq$ 0.13 mag, while the mean E($B-V$) value in the region within 15$^{\circ}$ of the Galactic plane is almost 1 mag. As a result, even if there were AGNs behind the high-extinction regions that are bright enough to be detected in \emph{Gaia}, they would tend to have brighter extinction-corrected \emph{Gaia} $G$ magnitudes than AGNs in the training set, and hence smaller $P_{\rm RF}$ values. The effective sky coverage is taken as the total area containing at least one AGN candidate from the R85 catalogue, which is approximately 36,000 deg$^2$. The average AGN number densities in the C75 and R85 catalogues are 76 deg$^{-2}$ and 61 deg$^{-2}$ respectively. 
Another clear feature in the spatial distributions of our AGN catalogues is the imprint of the \emph{Gaia} scanning law, i.e. the patchy or filamentary structures in Figure~\ref{fig:spatial_distribution}. As explained in Section~\ref{sect:completeness}, the \emph{Gaia} limiting magnitude is deeper in regions that have more repeated \emph{Gaia} observations. As a result, the catalogue completeness and hence AGN density distribution show correlations with the \emph{Gaia} scanning law. We expect this to improve in later \emph{Gaia} data releases. 

The top two panels in Figure~\ref{fig:distributions} show the normalised histograms of parallax and overall proper motion for the C75 (solid red) and R85 (dashed blue) AGN catalogues. AGN candidates with null parallaxes or proper motions are not included in the histograms. The two catalogues have very similar parallax distributions. Ignoring AGN candidates with null parallaxes, the mean and median parallax of the C75 (R85) catalogue are $-0.019$ ($-0.026$) mas and $-0.022$ ($-0.026$) mas. The mean parallax of the more reliable R85 catalogue is consistent with the global parallax zero point of $-0.029$ mas found for \emph{Gaia} DR2, considering the typical parallax uncertainty of 0.03--0.7 mas \citep{Lindegren18}. 

The bottom left panel in Figure~\ref{fig:distributions} shows the normalised, extinction-corrected \emph{Gaia} $G$-band magnitude distributions for the C75 (solid red) and R85 (dashed blue) AGN catalogues. At the faint end, the distributions for both samples drop sharply beyond $G \sim 20.6$ mag. We find that the C75 catalogue has a larger fraction of objects in faint magnitude bins compared to the R85 catalogue, implying that the contamination rate in the C75 catalogue becomes higher in fainter magnitude bins. 

\begin{figure*}
    \centering
    \includegraphics[width=0.8\textwidth]{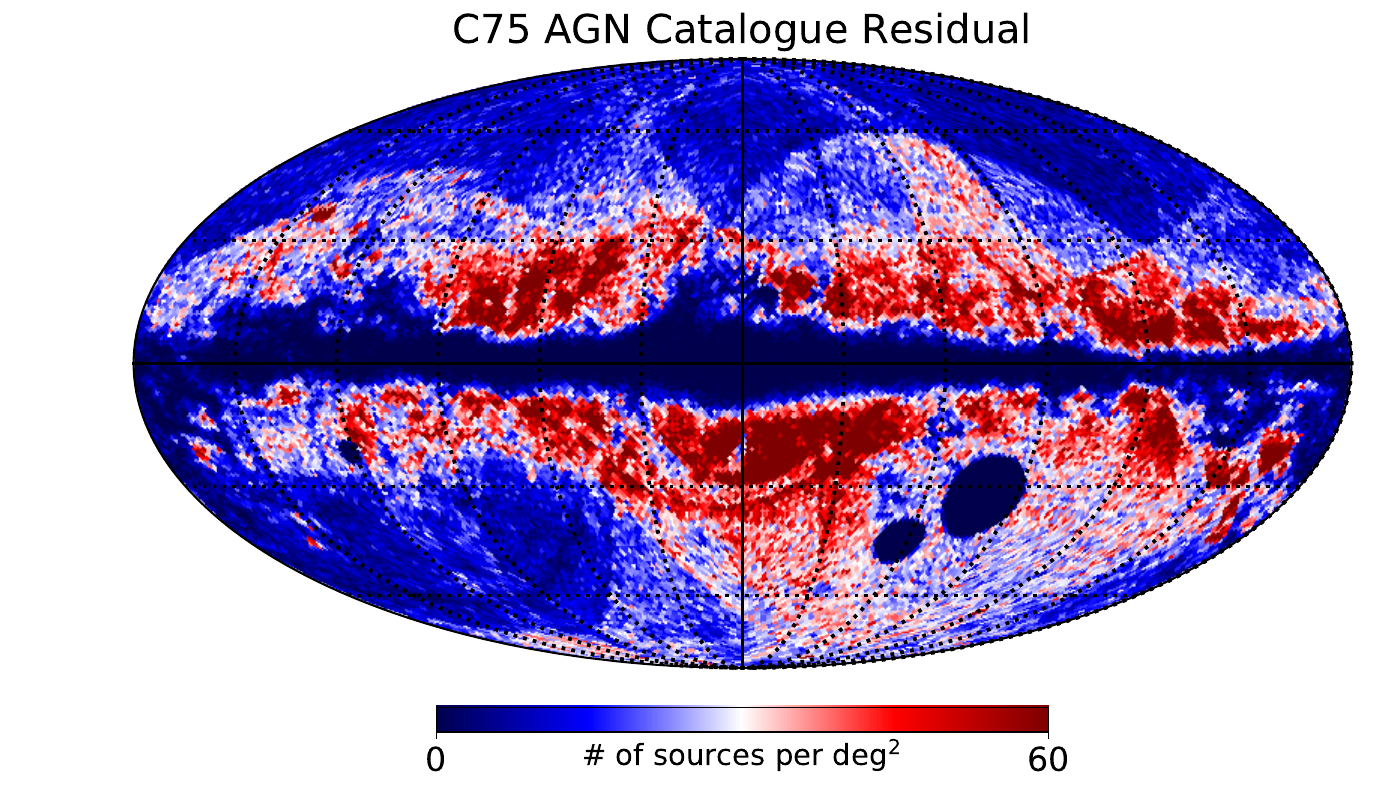}
    \includegraphics[width=0.8\textwidth]{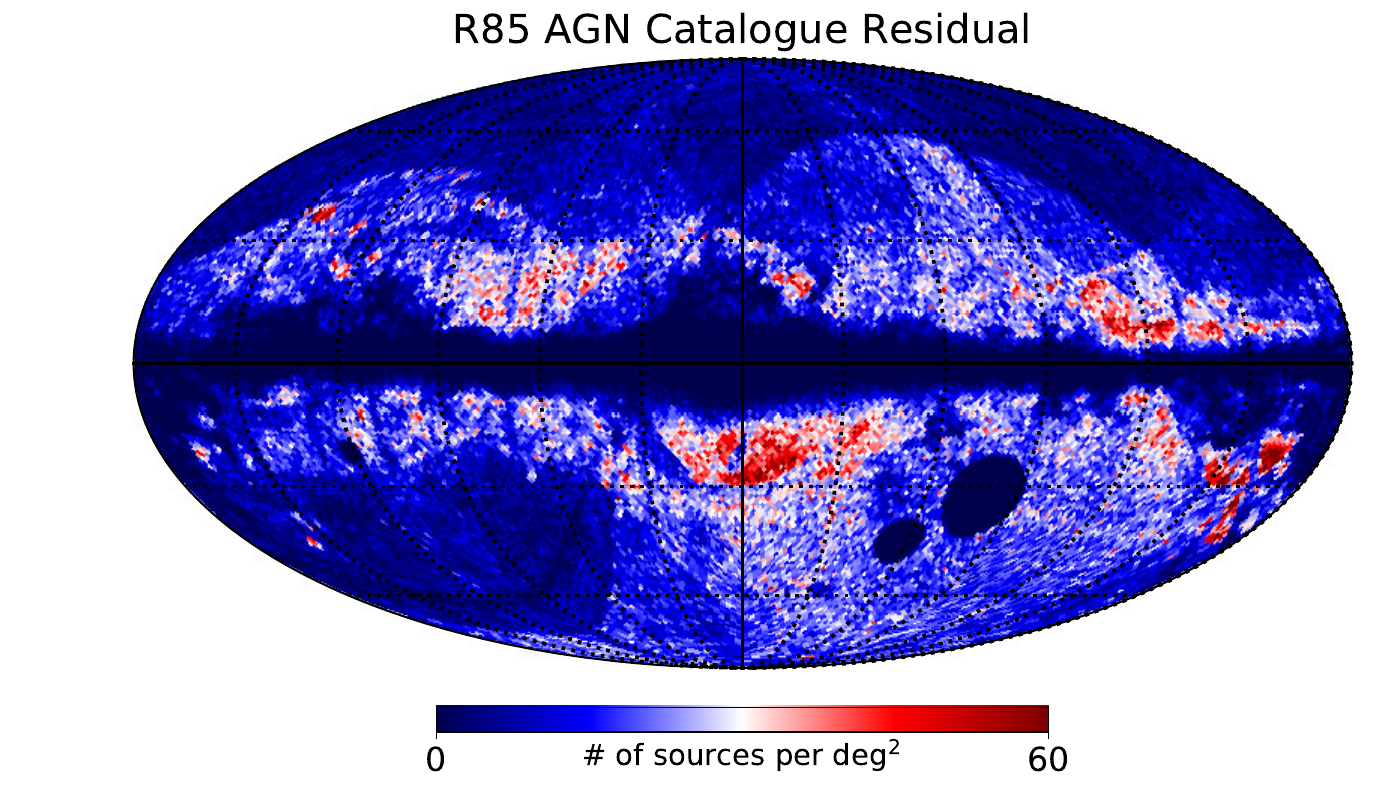}
    \caption{\label{fig:residual_distribution} Spatial distribution (in Mollweide projection) of new AGN candidates in the C75 and R85 AGN catalogues in the Galactic coordinate system after removing overlaps with the known AGN compilation.}
\end{figure*}

We apply the best-trained photometric redshift estimator to the C75 catalogue, and the bottom right panel in Figure~\ref{fig:distributions} shows the normalised histograms of the estimated redshifts for AGN candidates in the C75 (solid red) and R85 (dashed blue) catalogues. 76,620 (28,929) AGN candidates in the C75 (R85) catalogue are predicted to be at $z_{\rm phot} \geq 3$, and 1,602 (193) AGN candidates in the C75 (R85) catalogue are predicted to be at $z_{\rm phot} \geq 4$. Considering the photometric-redshift bias found using the test set, we expect the number of high-redshift ($z \gtrsim 3$) AGNs in our catalogues being higher than suggested by the estimated redshifts. 

\section{Discussions}
\label{sect:discussions}

\subsection{Comparisons with other AGN catalogues}

\begin{figure*}
    \centering
    \includegraphics[width=0.98\textwidth]{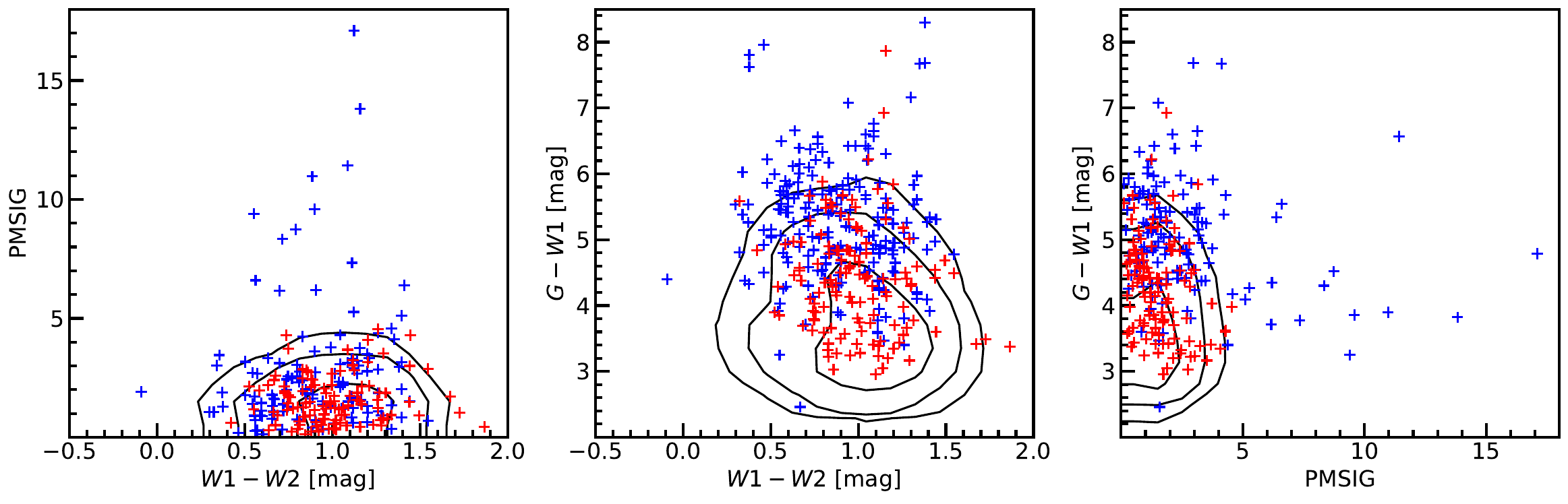}
    \caption{\label{fig:lensed_qsos_recovery} Distributions of the top three most important features, W1$-$W2, PMSIG, and G$-$W1, for the 333 known lensed quasar images in the \emph{Gaia}-unWISE sample. The black contours correspond to distributions of AGNs in the training set. Compared to AGNs in the training set or lensed quasar images recovered in the C75 catalogue (red symbols), lensed quasar images that are not in the C75 catalogue (blue symbols) tend to have smaller W1$-$W2 and larger PMSIG and G$-$W1.}
\end{figure*}

We build the AGN training set from the DR14Q catalogue because it is the largest spectroscopically confirmed AGN sample to date. To examine whether our RF classifier inherits any selection bias from this choice of training set, we compare our AGN catalogues with some known, large AGN catalogues selected in different wavelength domains and by various techniques in the literature. 

The MILLIQUAS catalogue \citep[version 5.7,][]{Flesch15} is a compendium of almost 2 million AGNs and high-confidence AGN candidates including the DR14Q sample, the 2-degree Field QSO sample \citep[2QZ,][]{Croom04}, QSO catalogues from the Large Sky Area Multi-object Fiber Spectroscopic Telescope \citep[LAMOSTQ,][]{Ai16, Dong18, Yao19}, the NBCKDE v3 catalogue \citep[NBCKv3,][]{Richards15}, the SDSS-XDQSO catalogue \citep[XDQSO,][]{Bovy11}, the AllWISE AGN catalogue \citep[WISEA,][]{Secrest15}, the Million Optical-Radio/X-ray Associations Catalogue \citep[MORX,][]{Flesch16}, with the remaining from various other discovery papers\footnote{ A complete list of the MILLIQUAS input catalogues and references can be found at \url{https://heasarc.gsfc.nasa.gov/W3Browse/all/milliquas.html}}. Cross-matching the MILLIQUAS catalogue with the \emph{Gaia}-unWISE sample using a matching radius of 0\farcs5 results in 1,166,573 matches, which are referred to as the MILLIQUAS-Gaia-unWISE sample. We find that 94.7\% and 89.4\% of the MILLIQUAS-Gaia-unWISE sample are successfully recovered in our C75 and R85 catalogues. We note that these recovery rates should not be directly compared to the completeness levels of the C75 and R85 catalogues because the MILLIQUAS-Gaia-unWISE sample is
not complete in the first place. Instead, the overall, high recovery rates demonstrate the effectiveness of our RF classifier. 

Breaking the MILLIQUAS-Gaia-unWISE sample apart, we find that the recovery rates for the DR14Q, 2QZ, and LAMOSTQ samples are higher than the above overall rates, at $\approx$98\% (C75) and $\approx$95\% (R85) respectively. The bulk of the DR14Q sample are used in the training process, so its recovery rates are expected to be higher than average. The similarly high recovery rates for the 2QZ and LAMOSTQ samples may be attributed to their target selections being similar to what are used for the DR14Q sample. The 2QZ quasars are selected based on optical $ub_{\rm J}r$ colours \citep{Smith05}, which is similar to how some of the SDSS DR7 quasars (a subset of the DR14Q sample) are selected. 
The LAMOSTQ sample is primarily selected using optical-infrared colours \citep{Wu10, Wu12, Ai16} together with the extreme deconvolution \citep{Bovy11} and kernel density estimation \citep{Richards09} techniques. The CORE sample in the DR14Q is selected based on the extreme deconvolution technique, and a part of the BONUS sample in the SDSS DR12 QSO catalogue (a subset of the DR14Q sample) is selected based on the extreme deconvolution and kernel density estimation techniques. 
For the MORX sample, the recovery rates are 82\% (C75) and 71\% (R85), significantly lower than the overall rates. The MORX sample included in the MILLIQUAS catalogue corresponds to AGNs that are discovered in radio/X-ray \citep{Flesch16}. Considering that radio/X-ray observations are less affected by dust obscuration compared to optical, the lower-than-average recovery rates for the MORX sample may indicate that our RF classifier is less efficient in selecting obscured AGNs. It is also possible that the MORX sample has a higher contribution from host galaxy emission which would result in redder $G-W1$ and bluer $W1-W2$ colours compared to AGNs in the training set \citep[e.g.,][]{Ostrovski17, Lemon19}. 

To determine the number of new AGN candidates in our catalogues, we cross-match the C75 and R85 catalogues with the known AGN compilation using an aggressive matching radius of 5$^{\prime \prime}$. We find that at least 911,622 and 515,246 AGN candidates in our C75 and R85 catalogues are previously unknown. Figure~\ref{fig:residual_distribution} shows the spatial distributions of these new AGN candidates, which we refer to as residual maps. Within the extensively observed and studied SDSS footprint, there are few new AGN candidates because our catalogues are limited by the \emph{Gaia} detection limit, which is brighter than those of the known AGN catalogues in this field. Although there have been searches for AGNs outside the SDSS footprint (mostly using the all-sky \emph{WISE} data), our AGN catalogues still find, on average, 30--50 new AGN candidates per deg$^2$ in those regions, demonstrating the high completeness of our AGN selection technique (e.g. Table~\ref{tb:performance}). Comparing the residual maps of the C75 and R85 catalogues, we find that the number densities of the low-probability AGN candidates close to the Galactic plane and bulge are higher than average, which we think is due to the higher overall source densities therein. 

Lastly, we examine how many known strongly-lensed quasars are recovered in our AGN catalogues. To date there are 204 known strongly-lensed quasar systems according to the Gravitationally Lensed Quasar Database\footnote{\url{https://www.ast.cam.ac.uk/ioa/research/lensedquasars/}} \citep{Lemon19}. In total, 333 lensed quasar images in 168 
known systems are in the \emph{Gaia}-unWISE sample, of which 126 lensed quasar images in 104 systems have large enough $P_{\rm RF}$ values to be included in the C75 catalogue. The recovery rate is much lower than found above for AGNs in general. Figure~\ref{fig:lensed_qsos_recovery} shows the top three most important features, W1$-$W2, PMSIG, and G$-$W1, for the 333 known lensed quasar images. We can see that the un-recovered lensed quasar images (blue symbols) generally have smaller W1$-$W2 and larger PMSIG and G$-$W1 than the recovered lensed quasar images (red symbols) or AGNs in the training set (black contours). From imaging data, we find that those un-recovered lensed quasar images are usually close to the lensing galaxies or clustered within small separations. They have lower $P_{\rm RF}$ values because 1) their $W1-W2$ and $G-W1$ colours are contaminated by the nearby lensing galaxies \citep[e.g.,][]{Lemon19}; 2) their proper motions and parallaxes are inaccurately inferred, perhaps due to \emph{Gaia} mis-assigning nearby images at each epoch; 3) they generally have CNT4 $> 1$, which makes them less similar to AGNs in our training set where more than 99\% of AGNs have CNT4 $=1$. We note that finding highly-clustered AGNs on small scales ($\lesssim$10$^{\prime \prime}$) in the presence of nearby, bright galaxies is essentially a different task from building a large and clean sample of AGNs, and a separate classifier/approach might be needed. 

\subsection{A wide-separation, strongly-lensed AGN candidate}

\begin{figure*}
    \centering
    \includegraphics[width=0.45\textwidth]{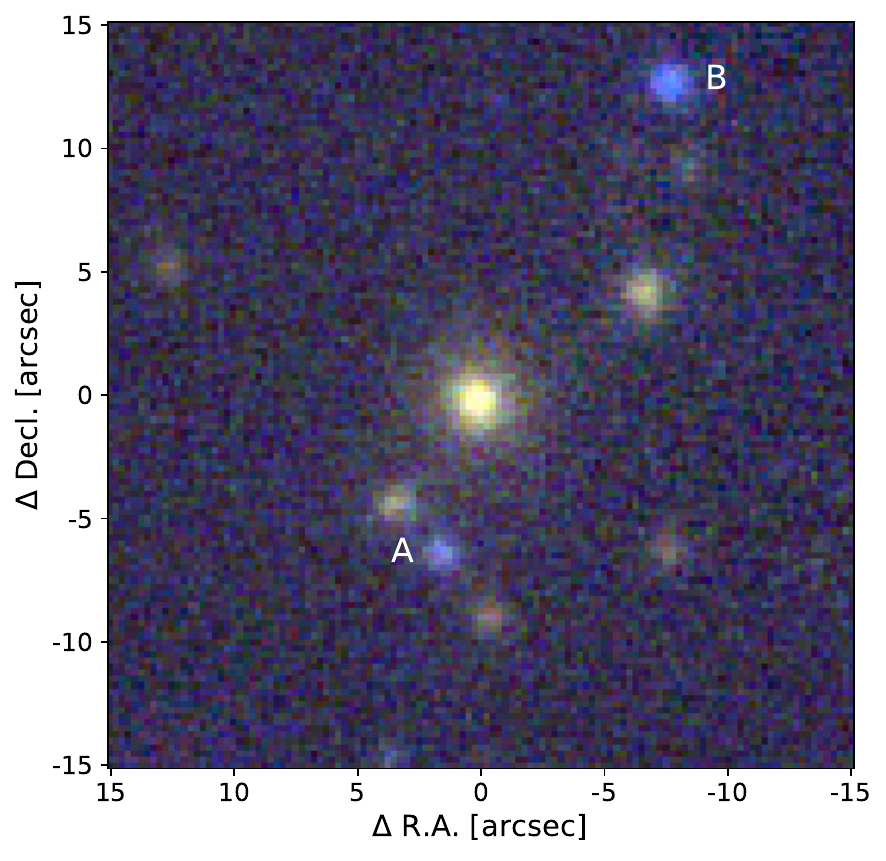}
    \includegraphics[width=0.53\textwidth]{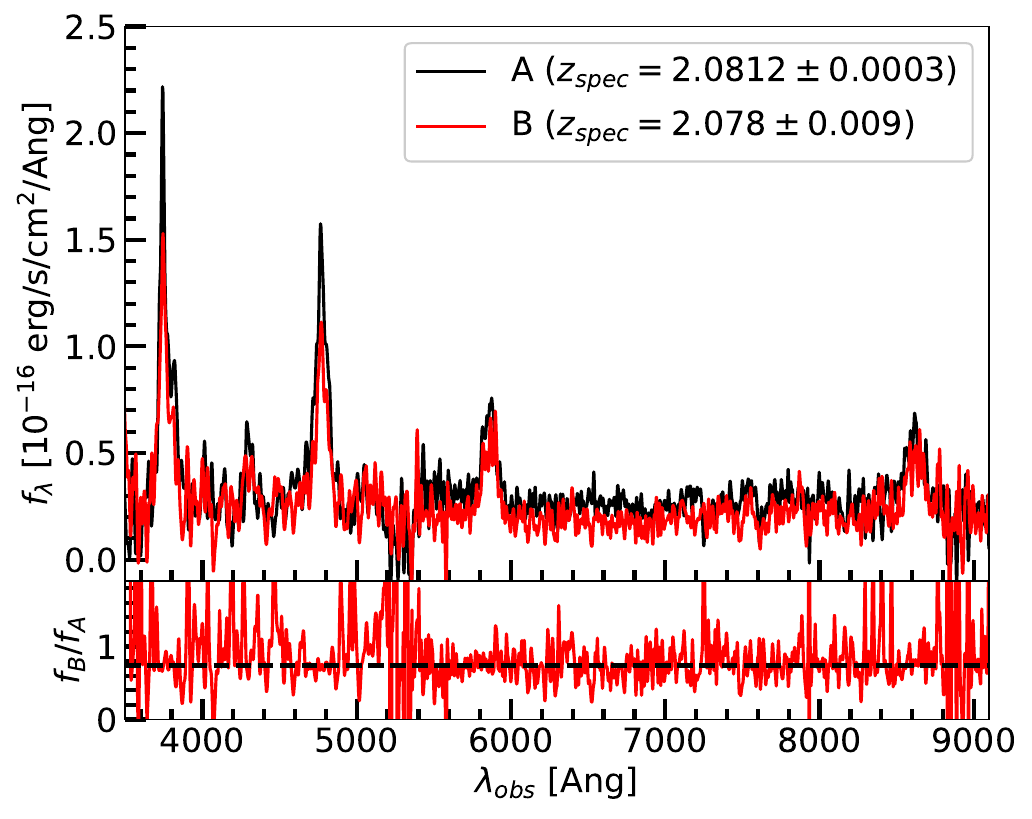}
    \caption{\label{fig:J1326} \emph{Left}: PanSTARRS imaging data of the new wide-separation strongly-lensed quasar SDSS\,J1326$+$4806. The bright object in the center is a BCG at $z=0.396$. Object A is a spectroscopically confirmed quasar at $z_A=2.0812 \pm 0.0003$. Object B is classified as an AGN in our catalogue with $P_{\rm RF}= 0.93$. The separation between A and B is 21\farcs06. \emph{Top right}: Smoothed WHT spectra of A (black) and B (red). Fitting the spectrum of B confirms it to be a quasar at $z_B =2.078 \pm 0.009$. \emph{Bottom right}: Flux ratio of B to A. The median flux ratio is 0.74, as indicated by the black dashed line. }
\end{figure*}

Although our AGN catalogues are not effective in finding small-separation, strongly-lensed AGN systems, they are useful in finding wide-separation strong-lens systems. It has been shown that strongly-lensed AGNs with wide image separations ($> 10^{\prime \prime}$) are valuable cosmological probes \citep[e.g.,][]{Narayan88, Turner90, Fukugita90, Wambsganss95, Kochanek95b, Kochanek96, Lopes04, Oguri04, Li07, Oguri12b}. However, only four known strongly-lensed AGNs have maximum image separations larger than $> 10^{\prime \prime}$ \citep[][]{Inada03, Inada06, Dahle13, Shu18}. 
We thus carry out a search for wide-separation, strongly-lensed AGNs by identifying brightest cluster galaxies (BCGs) that have at least two AGN candidates from our C75 catalogue located within a circular aperture of 30$^{\prime \prime}$ radius. The BCG sample we use is compiled from \citet{Wen11, Wen15, Wen18a, Wen18b}, which contains 209,419 BCGs (duplicates not removed) up to redshift of one. 57 unique BCGs with at least two neighbouring AGN candidates are found, and their optical images are visually inspected. We re-discover two previously known wide-separation, strongly-lensed quasar systems SDSS\,J1004$+$4112 \citep{Inada03} and SDSS\,J1029$+$2623 \citep{Inada06}. The other two known wide-separation, strongly-lensed quasar systems, SDSS\,J0909$+$4449 \citep{Shu18} and SDSS\,J2222$+$2745 \citep{Dahle13}, are not recovered because they only have zero and one lensed quasar image detected in \emph{Gaia} DR2. In addition, we identify a high-probability strongly-lensed AGN candidate --- SDSS\,J1326$+$4806. The majority of the rest of the BCGs have AGN candidates with significantly different optical colours, and therefore unlikely to be images of the same AGN, or the BCG does not lie between the AGN candidates.

The left panel in Figure~\ref{fig:J1326} shows a colour cutout centered on the BCG of SDSS\,J1326$+$4806 made from $gri$ imaging data from the Panoramic Survey Telescope and Rapid Response System (PanSTARRS) survey \citep{Chambers16}. The BCG, at R.A.=201.50006$^{\circ}$, Decl.=48.11208$^{\circ}$, is an SDSS spectroscopically-confirmed massive early-type galaxy at $z=0.396$. Two blue, point-like sources, labeled as A and B, are located on either side of the BCG, consistent with the image configuration of a doubly lensed system. The separation between A and B is 21\farcs06. Our AGN classifier suggests that A and B are very likely to be AGNs with $P_{\rm RF} \, (\rm A) = 0.99$ and $P_{\rm RF} \, (\rm B) = 0.93$. In fact, source A was spectroscopically confirmed to be a $z_A=2.0812 \pm 0.0003$ AGN by the Baryon Oscillation Spectroscopic Survey \citep{Bolton12}. 

To determine the nature and redshift of B, we obtained low-resolution spectra for A and B with the Intermediate-dispersion Spectrograph and Imaging System on the William Herschel Telescope (WHT) on the night of February 11, 2019. The R158R (1.81 \AA \ pixel$^{-1}$) and R300B (0.86 \AA \ pixel$^{-1}$) gratings were used on the red and blue arms, respectively, along with the standard 5300 \AA \ dichroic and GG495 second-order cut filter in the red arm. The right panel in Figure~\ref{fig:J1326} shows the smoothed, reduced spectra for A (black) and B (red), which confirms that B is indeed an AGN with a spectral profile that appears to be similar to A. Fitting the spectrum of B using a linear combination of quasar eigenspectra following \citet{Bolton12} further suggests $z_B=2.078 \pm 0.009$, consistent with the spectroscopic redshift of A. 

Both A and B have experienced substantial variations in brightness over the past $\sim 16$ years. The SDSS data in 2003 showed that the $g$-band AB magnitude of A and B were about 21 mag and 22 mag respectively, with A being brighter than B. The multi-epoch photometry from PanSTARRS DR2 taken between the year of 2011 and 2014 showed significant brightness variations, with the largest change reaching more than 1 magnitude. In particular, B was brighter than A when averaging over the PanSTARRS period, as indicated in the left panel of Figure~\ref{fig:J1326}. The PanSTARRS $g$-band mean AB magnitude of A and B were about 21.6 mag and 21 mag respectively. The median flux ratio of B to A from recent WHT spectroscopic data is 0.74, indicating that A now has become brighter than B again. Nevertheless, no clear correlation between brightness variations in A and B is detected. 

We consider a simple lens model for SDSS\,J1326$+$4806 consisting of a singular isothermal sphere (SIS) mass distribution in an external shear field. The total number of free parameters is 7 (assuming the SIS mass component and the external shear field are co-centred). Considering the substantial brightness variations in A and B, we only use the relative positions of the BCG, A, and B as constraints, but not the flux ratios between A and B. As a result, the number of free parameters is more than the number of constraints, and no unique lens model can be determined. Nevertheless, the goal of this procedure is to examine whether the image configuration of SDSS\,J1326$+$4806 can be explained by a typical lens model with reasonable parameters. We optimize the model parameters with the {\tt lensmodel} toolkit \citep{Keeton01}, and find that the relative positions can be perfectly recovered (as expected for this under-constrained problem). All the model parameters have reasonable values. The best-fit Einstein radius of the SIS component is 10\farcs3, consistent with the 21\farcs06 separation between A and B. It suggests that the total projected mass within the Einstein radius is $\approx 2.1 \times 10^{13} M_{\odot}$. 
On the other hand, \citet{Wen12} estimated the $r_{200}$ radius of this cluster to be 1.51 Mpc. Assuming that the dark matter distribution of this cluster follows a simple Navarro-Frenk-White (NFW) profile \citep{Navarro96, Navarro97}, the total dark-matter mass within the sphere of radius $r_{200}$ is approximately $M_{200} = 5.6 \times 10^{14} M_{\odot}$. The typical concentration for dark-matter halos of this mass scale at $z \sim 0.4$ is about 5 \citep[e.g.,][]{Duffy08, Maccio08, Zhao09, Klypin11, Prada12, Auger13, Diemer15}. 
The total projected dark-matter mass within the Einstein radius (57 kpc in physical unit) is thus $2.0 \times 10^{13} M_{\odot}$, in close agreement with the required mass by strong gravitational lensing. 

Based on the analyses above, SDSS\,J1326$+$4806 has a very high probability of being a strongly-lensed AGN. Follow-up higher-resolution spectroscopic and deeper imaging data could pin down the lensing nature of this system. If confirmed, SDSS\,J1326$+$4806 will be the second most widely-separated strongly-lensed AGN discovered so far. More wide-separation, strongly-lensed AGN systems are expected to be discovered by cross-matching the C75 AGN catalogue with other catalogues of galaxy groups and clusters. 

\subsection{Future Prospects}

It is worth mentioning that as more repeated \emph{Gaia} observations will be conducted in the coming years, we expect the overall limiting magnitude of future \emph{Gaia} data releases to become similar to the current value of the Bo\"{o}tes field or even deeper in some regions. Considering that in the Bo\"{o}tes reference field, the current completeness at the C75 threshold is 93.6\% and the reliability at the R85 threshold is 90.7\%, we expect the quality of AGN catalogues built from future \emph{Gaia} data releases to improve substantially both in completeness and reliability. In addition, the sample size and quality in astrometry and photometry of future \emph{Gaia} data releases are also expected to improve with beneficial effects for future AGN catalogues. 

On average, \emph{Gaia} will measure astrometrically each of its targets $\sim 70$ times over the nominal five-year operation period since 2013, and 10 photometric measurements in the $G$ band are made during each astrometric measurement \citep{Prusti16}. In total, every \emph{Gaia} source will therefore have $\sim 700$ $G$-band measurements in five years. In \emph{Gaia} DR2 (data from the first 22 months of operation), the average and highest number of $G$-band measurements for AGNs in the \emph{Gaia}-unWISE-DR14 QSO sample is 211 and 1100 respectively. However, \emph{Gaia} will not release the multi-epoch photometric data until the end of the mission, at which point all the AGN candidates in our catalogues will have \emph{Gaia} light curves spanning a time scale of five years. These light curves will be helpful in identifying variable AGNs and even optical changing-look AGNs. These are AGNs that show optical spectral feature transitions involving appearance and disappearance of broad emission lines on time scales of years or decades. There are a few tens of known optical changing-look AGNs so far \citep[e.g.,][]{Denney14, LaMassa15, Ruan16, MacLeod16, Gezari17, Yang18, WangJ18}. The physical mechanisms responsible for the transitions are still not fully understood. A large sample of variable AGNs and changing-look AGNs with a wide range of properties including redshift, luminosity, and black hole mass can help to better understand the structure of the accretion disc and broad line region and the evolution of AGNs. 
Our AGN catalogues, which include AGNs up to redshift $\sim 4$, can be a useful input catalogue for future spectroscopic surveys that study AGNs and large scale structures, especially ones in the southern hemisphere, for example, 4MOST \citep{deJong19, Merloni19, Richard19}. 

\section{Conclusion}
\label{sect:conclusion}

In this work, we perform an AGN/non-AGN classification of more than 641 million sources in the \emph{Gaia}-unWISE sample across the entire sky using astrometric and photometric data from the latest data releases of \emph{Gaia} and \emph{WISE}. We use the supervised machine learning algorithm random forest (RF) to estimate the probability of a source being an AGN, $P_{\rm RF}$. 
We construct two AGN catalogues, C75 and R85, by applying two different $P_{\rm RF}$ threshold cuts that deliver an overall completeness of 75\% ($\approx 90\%$ at $G \leq 20$ mag) and an overall reliability of 85\% respectively. The C75 catalogue contains 2,734,464 AGN candidates with $P_{\rm RF} \geq 0.69$, of which 2,182,193 AGN candidates with $P_{\rm RF} \geq 0.94$ comprise the R85 catalogue (Figure~\ref{fig:spatial_distribution}). We estimate the photometric redshifts of the AGN candidates using a RF regressor. We find that 76,620 and 1,602 AGN candidates in the C75 catalogue are predicted to be at redshifts higher than 3 and 4 respectively. 

Comparing to \emph{WISE}-only AGN selection techniques used in \citet{Stern12} and \citet{Assef18}, our RF classifier using both optical and mid-IR data achieves significantly better true positive and false positive rates when applied to the \emph{Gaia}-unWISE sample (see Table~\ref{tb:performance}). Among the 1,166,573 known AGNs and high-confidence AGN candidates in the MILLIQUAS that are also catalogued in the \emph{Gaia}-unWISE sample, 94.7\% and 89.4\% are successfully recovered in our C75 and R85 catalogues. Cross-matching against the known AGN compilation including almost 29 million AGNs and AGN candidates with an aggressive matching radius of 5$^{\prime \prime}$, we find that at least $\approx$0.91 (0.52) million AGN candidates in our C75 (R85) catalogue are new discoveries. 

The large sample of AGN candidates provided in this work is a useful resource for many applications. As an example, we have identified a strongly-lensed AGN candidate, SDSS\,J1326$+$4806, with an image separation of 21\farcs06 by cross-matching the C75 catalogue with a sample of known brightest cluster galaxies or BCGs (Figure~\ref{fig:J1326}). The BCG in SDSS\,J1326$+$4806 is at $z=0.396$, and the two AGN candidates on either side of the BCG are spectroscopically confirmed to be true AGNs at $z \sim 2.08$ with similar spectral profiles. A simple singular isothermal sphere plus external shear lens model can explain the relative positions between the BCG and the two AGNs. The total mass within the inferred Einstein radius required by strong gravitational lensing is in close agreement with the mass of dark matter within the same aperture when assuming dark matter in SDSS\,J1326$+$4806 following a simple NFW profile. Follow-up imaging and spectroscopic data will pin down the lensing nature of this system. 

Moreover, all the AGN candidates in our catalogue will eventually have light curves consisting of, on average, $\sim$70-epoch photometry across five years from \emph{Gaia}, which are very helpful for identifying highly-variable AGNs and changing-look AGNs. Our AGN catalogues are also useful for future spectroscopic surveys such as 4MOST.  

\section*{Acknowledgements}

We thank Qiusheng Gu, Paul Hewett, George Lansbury, Peter McGill, Leigh Smith, and Zhonglue Wen for helpful discussions. We thank the anonymous referee for a thoughtful report. Y.S. has been supported by the Royal Society -- K.C. Wong International Fellowship (NF170995).
SK is partially supported by NSF grant AST-1813881, Heising-Simon's foundation grant 2018-1030.
This work made used of the Whole Sky Database (wsdb) created by Sergey Koposov and maintained at the Institute of Astronomy, Cambridge by Sergey Koposov, Vasily Belokurov and Wyn Evans with financial support from the Science \& Technology Facilities Council (STFC) and the European Research Council (ERC). This software made use of the Q3C software \citep{koposov06}.
This work has made use of data from the European Space Agency (ESA) mission {\it Gaia} (\url{https://www.cosmos.esa.int/gaia}), processed by the {\it Gaia} Data Processing and Analysis Consortium (DPAC,\url{https://www.cosmos.esa.int/web/gaia/dpac/consortium}). Funding for the DPAC has been provided by national institutions, in particular the institutions participating in the {\it Gaia} Multilateral Agreement. This publication makes use of data products from the \emph{Wide-field Infrared Survey Explorer}, which is a joint project of the University of California, Los Angeles, and the Jet Propulsion Laboratory/California Institute of Technology, and NEOWISE, which is a project of the Jet Propulsion Laboratory/ California Institute of Technology. \emph{WISE} and NEOWISE are funded by the National Aeronautics and Space Administration.




\bibliographystyle{mnras}
\bibliography{ref}



\appendix

\section{unWISE Completeness and Limiting Magnitude Maps}

\begin{figure*}
    \centering
    \includegraphics[width=0.48\textwidth]{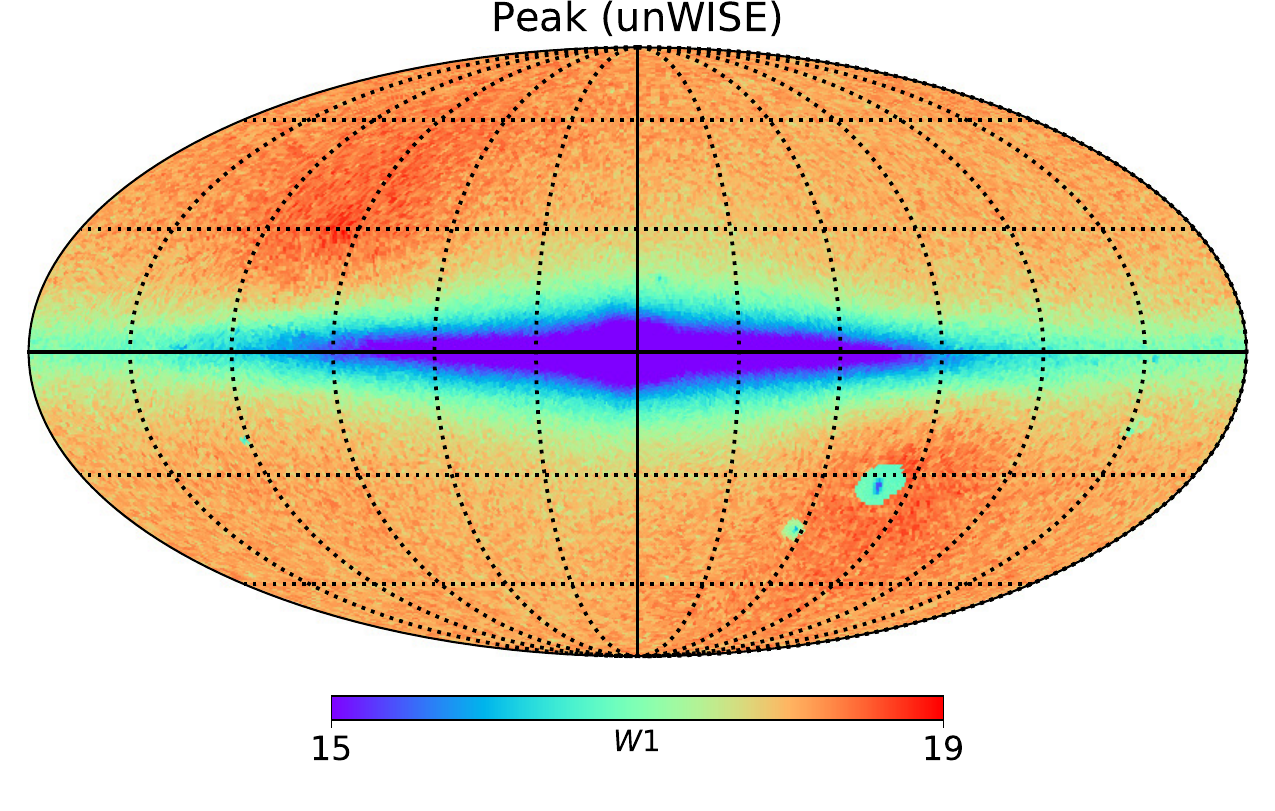}
    \includegraphics[width=0.48\textwidth]{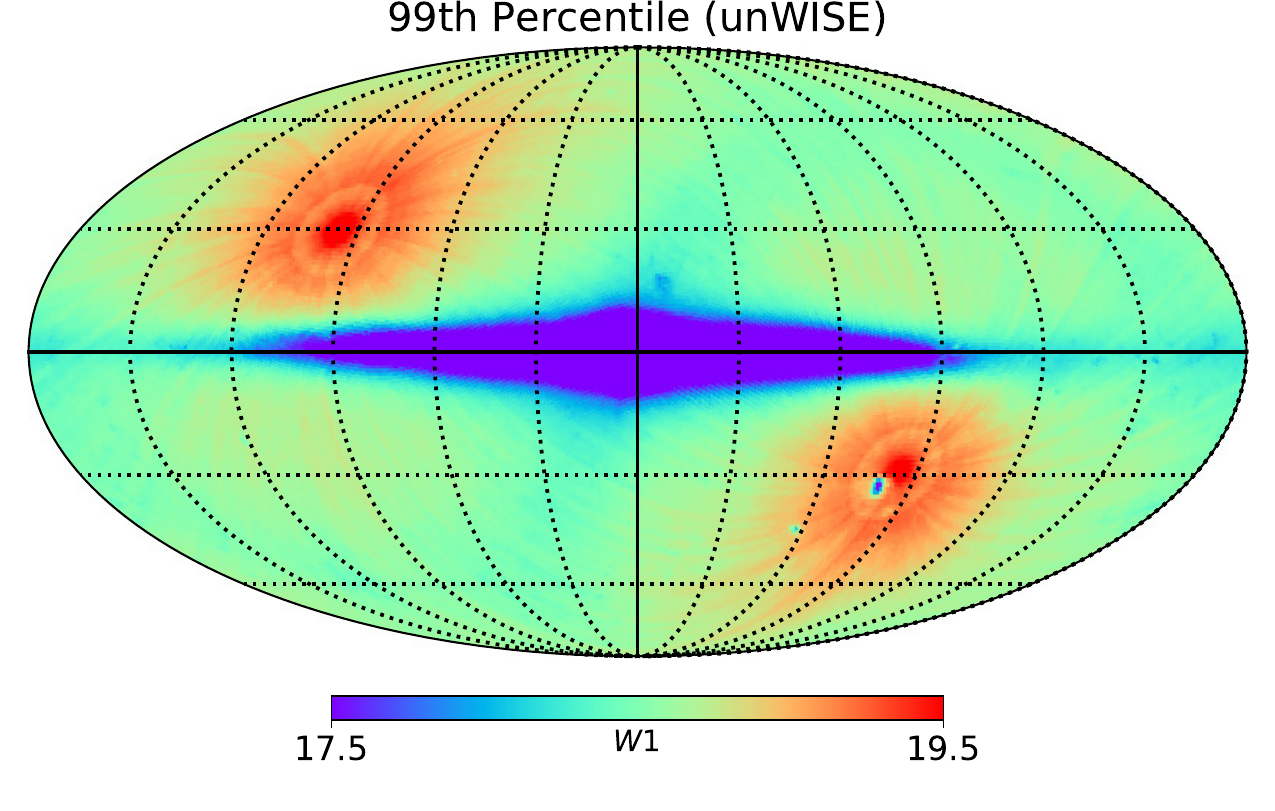}
    \includegraphics[width=0.42\textwidth]{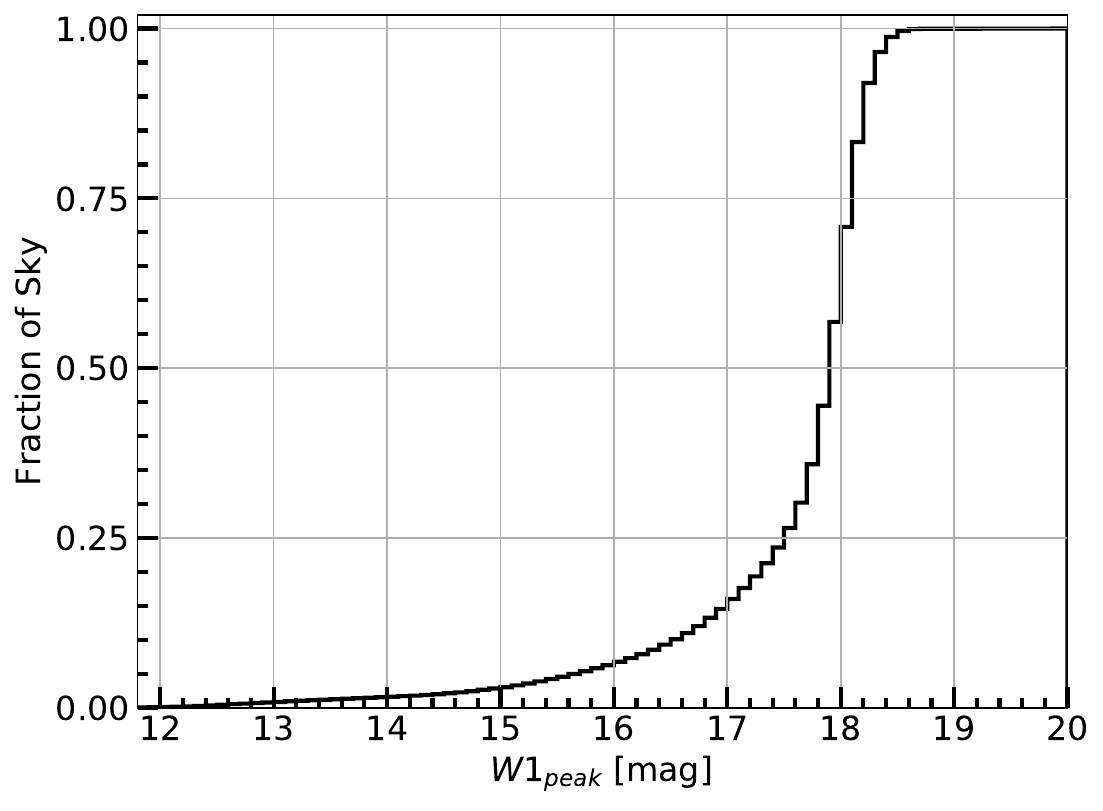}
    \hspace{0.8cm}
    \includegraphics[width=0.42\textwidth]{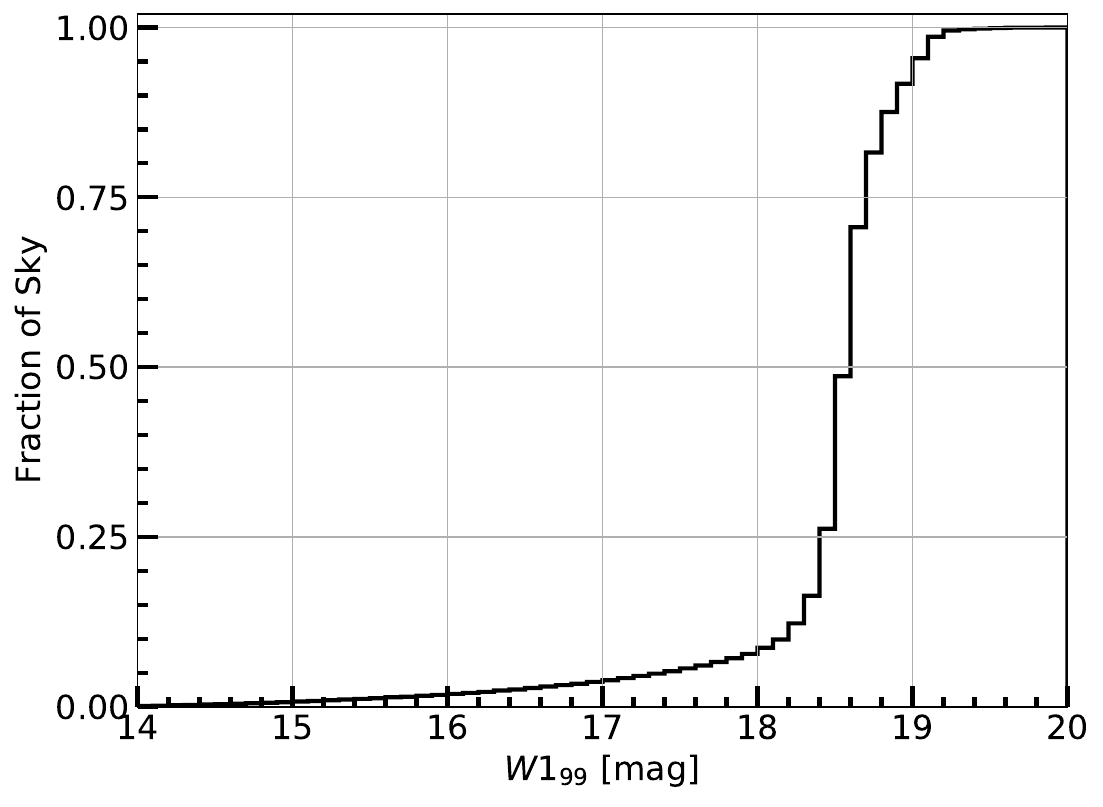}
    \caption{\label{fig:unWISE_W1_Limit_Distribution} The same set of plots as Figure~\ref{fig:Gaia_G_Limit_Distribution}, but for $W1$ for 2,094,307,508 unWISE sources with $W1 \geq 8$ mag. }
\end{figure*}

\begin{figure*}
    \centering
    \includegraphics[width=0.48\textwidth]{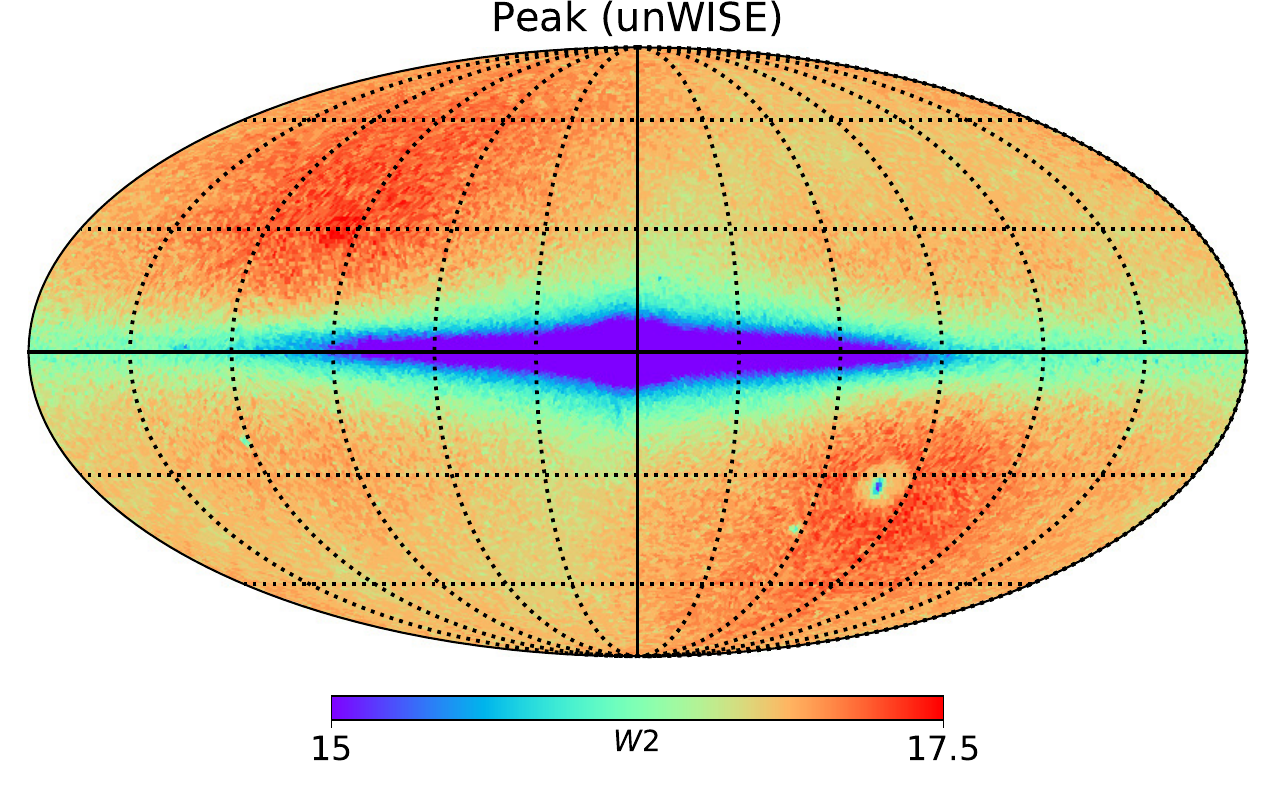}
    \includegraphics[width=0.48\textwidth]{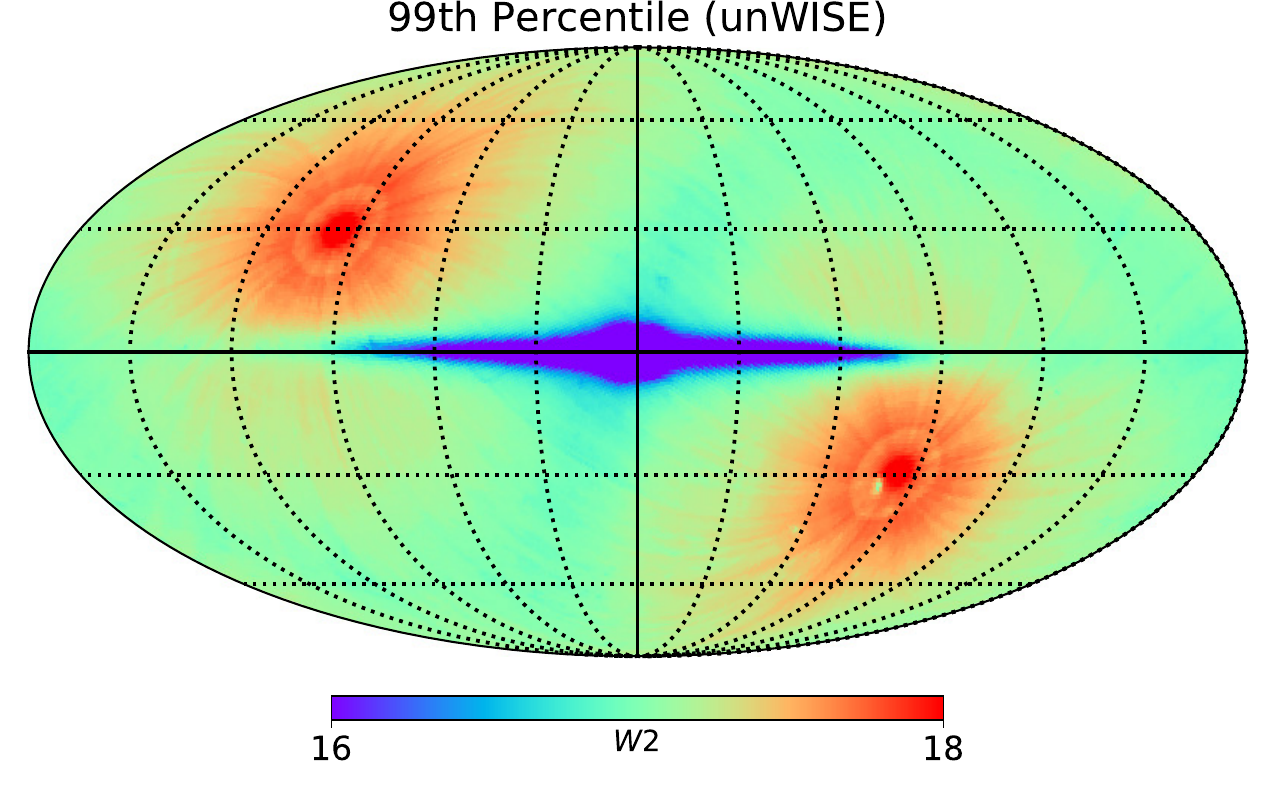}
    \includegraphics[width=0.42\textwidth]{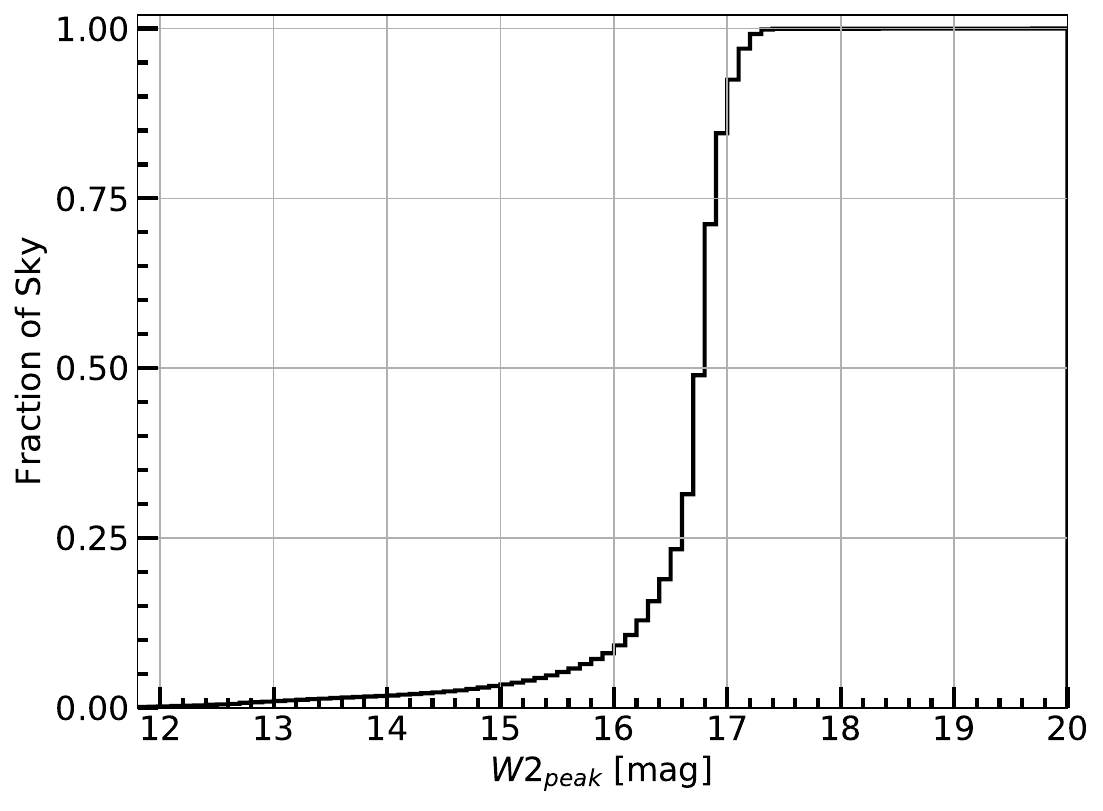}
    \hspace{0.8cm}
    \includegraphics[width=0.42\textwidth]{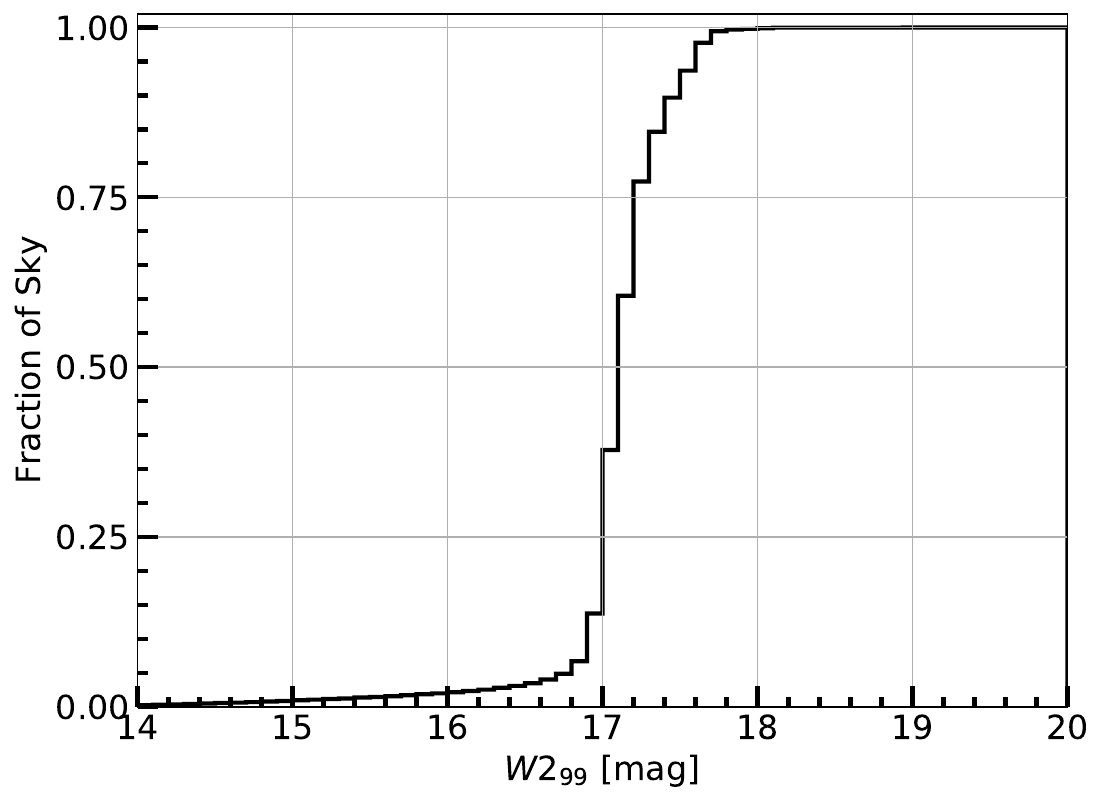}
    \caption{\label{fig:unWISE_W2_Limit_Distribution} The same set of plots as Figure~\ref{fig:Gaia_G_Limit_Distribution}, but for $W2$ for 1,180,720,229 unWISE sources with $W2 \geq 8$ mag.}
\end{figure*}

Figures~\ref{fig:unWISE_W1_Limit_Distribution} and \ref{fig:unWISE_W2_Limit_Distribution} show the spatial distributions and one-dimensional cumulative sky coverage histograms of $W1_{\rm peak}$, $W1_{\rm 99}$, $W2_{\rm peak}$, and $W2_{\rm 99}$ for the unWISE sub-samples. 

\section{Description of the catalogue}
\label{sect:catalogue_description}

The C75 AGN catalogue is publicly available as a FITS file at \url{https://www.ast.cam.ac.uk/~ypshu/AGN_Catalogues.html}. Descriptions of all the columns in the FITS file are summarised in Table~\ref{tb:catalogue_description}. The R85 AGN catalogue can be constructed from the C75 AGN catalogue by applying a probability threshold cut of $\text{PROB\_RF} \geq 0.94$.  

\begin{table*}
\centering
\caption{\label{tb:catalogue_description} Format of the AGN catalogue FITS file.}
\begin{tabular}{l l l}
\hline
\hline
Column & Name & Description \\
\hline
1 & RA & Right ascension in decimal degrees from \emph{Gaia} DR2 (J2015.5) \\
2 & DEC & Declination in decimal degrees from \emph{Gaia} DR2 (J2015.5) \\
3 & GAIA\_SOURCEID & Unique Gaia source identifier {\tt source\_id} \\
4 & UNWISE\_OBJID & Unique unWISE source identifier {\tt unwise\_objid} \\
5 & PLX & Parallax in milli-arcsec (mas) from \emph{Gaia} DR2, set to -999 if null \\
6 & PLX\_ERR & Error in parallax in mas from \emph{Gaia} DR2, set to -999 if null \\
7 & PMRA & Proper motion in right ascension direction (mas/year) from \emph{Gaia} DR2, set to -999 if null \\
8 & PMRA\_ERR & Error in proper motion in right ascension direction (mas/year) from \emph{Gaia} DR2, set to -999 if null \\
9 & PMDEC & Proper motion in declination direction (mas/year) from \emph{Gaia} DR2, set to -999 if null \\
10 & PMDEC\_ERR & Error in proper motion in declination direction (mas/year) from \emph{Gaia} DR2, set to -999 if null \\
11 & PLXSIG & Parallax significance defined as $\mid \frac{\text{parallax}}{\text{parallax\_error}} \mid$, set to -999 if null \\
12 & PMSIG & Proper motion significance defined as  $\sqrt{(\frac{\text{pmra}}{\text{pmra\_error}})^2+(\frac{\text{pmdec}}{\text{pmdec\_error}})^2}$, set to -999 if null \\
13 & EBV & Galactic E(B-V) reddening from \citet{Schlegel98} \\ 
14 & N\_OBS & Number of observations contributing to $G$ photometry \\ 
\hline
15 & G & \emph{Gaia} DR2 $G$-band mean magnitude (extinction corrected) \\
16 & BP & \emph{Gaia} DR2 BP-band mean magnitude (extinction corrected) \\
17 & RP & \emph{Gaia} DR2 RP-band mean magnitude (extinction corrected) \\
18 & W1 & unWISE $W1$-band magnitude \\
19 & W2 & unWISE $W2$-band magnitude \\
20 & BP\_G & \emph{Gaia} DR2 BP-$G$ colour (extinction corrected), set to 999 if null \\
21 & BP\_RP & \emph{Gaia} DR2 BP-RP colour (extinction corrected), set to 999 if null \\
22 & G\_RP & \emph{Gaia} DR2 $G$-RP colour (extinction corrected), set to 999 if null \\
23 & G\_W1 & \emph{Gaia} DR2 $G$ - unWISE $W$1 colour (extinction corrected) \\
24 & GW\_SEP & Separation (in arcsec) between a \emph{Gaia} source and its unWISE counterpart \\
25 & W1\_W2 & unWISE $W$1-$W$2 colour \\
\hline
26 & G\_VAR & Variation in \emph{Gaia} $G$-band flux defined as $\sqrt{{\tt PHOT\_G\_N\_OBS}} \times \frac{{\tt PHOT\_G\_MEAN\_FLUX\_ERROR}}{{\tt PHOT\_G\_MEAN\_FLUX}}$ \\
27 & BPRP\_EF & BP/RP excess factor from \emph{Gaia} DR2 ({\tt PHOT\_BP\_RP\_EXCESS\_FACTOR}) \\
28 & AEN & Astrometric excess noise from \emph{Gaia} DR2 ({\tt ASTROMETRIC\_EXCESS\_NOISE}) \\
29 & GOF & Goodness-of-fit statistic of the astrometric solution from \emph{Gaia} DR2 ({\tt ASTROMETRIC\_GOF\_AL}) \\
\hline
30 & CNT1 & Number of \emph{Gaia} DR2 sources within a 1$^{\prime \prime}$-radius circular aperture \\
31 & CNT2 & Number of \emph{Gaia} DR2 sources within a 2$^{\prime \prime}$-radius circular aperture \\
32 & CNT4 & Number of \emph{Gaia} DR2 sources within a 4$^{\prime \prime}$-radius circular aperture \\
33 & CNT8 & Number of \emph{Gaia} DR2 sources within a 8$^{\prime \prime}$-radius circular aperture \\
34 & CNT16 & Number of \emph{Gaia} DR2 sources within a 16$^{\prime \prime}$-radius circular aperture \\
35 & CNT32 & Number of \emph{Gaia} DR2 sources within a 32$^{\prime \prime}$-radius circular aperture \\
\hline
36 & PHOT\_Z & Photometric redshift \\
37 & PROB\_RF & AGN probability \\
\hline
\hline
\end{tabular}
\end{table*}

\bsp	
\label{lastpage}
\end{document}